\documentclass[letterpaper]{article}
\usepackage[utf8]{inputenc}
\usepackage{geometry}
\usepackage{listings}
\usepackage{amsmath}
\usepackage{amsfonts}
\usepackage{natbib}
\usepackage{hyperref}
\usepackage{afterpage}
\usepackage{rotating}
\usepackage{pdflscape}
\usepackage{authblk}
\usepackage{lscape} 
\usepackage[final]{pdfpages}
\usepackage{float}
\usepackage{changepage}
\usepackage[normalem]{ulem}
\usepackage{caption}
\usepackage{subcaption}
\usepackage{array}
\newcolumntype{L}{>{\centering\arraybackslash}m{3cm}}
\usepackage{xcolor}
\usepackage{tikz}
\usepackage{hyperref}
\hypersetup{ colorlinks=true, citecolor=blue, linkcolor=blue, urlcolor=blue}
\usepackage{xcolor}
\colorlet{linkequation}{blue}

\usepackage{xcolor}
\definecolor{OliveGreen}{cmyk}{0,0.05,0.05,0.05}

\usepackage{setspace}
\usepackage{pdflscape}
\usepackage{graphicx}
\usepackage[justification=justified]{caption}
\usepackage{afterpage}
\usepackage{multicol}
\usepackage{multirow}
\usepackage{soul}
\usepackage{lineno}
\usepackage{dashrule}
\usepackage[english]{babel}
\usepackage{wrapfig}
\usepackage{enumitem}
\usepackage{authblk}
\usepackage{siunitx,booktabs}
\begin{document}

\title{\textbf{Bayesian Hierarchical Models For Multi-type Survey Data Using  Spatially Correlated Covariates Measured With Error }}

\author[$\dagger$]{Saikat Nandy}
\author[1,2]{Scott H. Holan}
\author[3]{Jonathan R. Bradley}
\author[1]{Christopher K. Wikle}

\affil[1]{Department of Statistics, University of Missouri, Columbia, MO, USA}
\affil[2]{Office of the Associate Director for Research and Methodology, U.S. Census Bureau, DC, USA}
\affil[3]{Department of Statistics, Florida State University, Tallahassee, FL, USA}
\affil[$\dagger$]{Corresponding author: Department of Statistics, University of Missouri, Columbia, MO, USA; \href{mailto: snandy@mail.missouri.edu}{snandy@mail.missouri.edu}}
\renewcommand\Affilfont{\itshape\small}
\date{}

\maketitle
\begin{abstract}
We introduce Bayesian hierarchical models for predicting high-dimensional tabular survey data which can be distributed from one or multiple classes of distributions (e.g., Gaussian, Poisson, Binomial etc.). We adopt a Bayesian implementation of a Hierarchical Generalized Transformation (HGT) model to deal with the non-conjugacy of non-Gaussian data models when estimated using a Latent Gaussian Process (LGP) model. Survey data are usually prone to high degree of sampling error, and we use covariates that are prone to measurement error as well as those free of any such error. 
A classical measurement error component is defined to deal with the sampling error in the covariates. The proposed models can be high dimensional and we employ the notion of basis function expansions to provide an effective approach to dimension reduction. The HGT component lends flexibility to our model to incorporate multi-type response datasets under a unified latent process model framework. To demonstrate the 
applicability of our methodology, we provide the results from simulation studies and data applications arising from a dataset consisting of the U.S. Census Bureau’s American Community Survey (ACS) 5-year period estimates of the total population count under the poverty threshold and the ACS 5-year period estimates of median housing costs at the county level across multiple states in the U.S.A. 

\end{abstract}
\section{Introduction}

Sample surveys have long been popular and efficient means for collecting data on economic, labor, and health variables, measuring unemployment, determining income and poverty status, and other socio-demographic variables. 
These data are tabulated starting at the unit level, up to an aggregate level like county or state. While sample surveys are reliable sources for direct estimates of means or totals of variables of interest for large areas or domains, for smaller domain sample size plays a crucial role in prediction bias and in determining the uncertainty associated with these estimates. Domains may be defined by geographic areas, socio-demographic groups, or other sub-populations \citep{Rao2015}. 
A domain (area) is regarded as large if the domain-specific sample is large enough to yield ``direct estimates" with adequate precision whereas a small area would denote any domain for which direct estimates of adequate precision cannot be produced \citep{Rao2015}.

Failure to produce reliable estimates for small areas may occur for several reasons, among which the most common one would be the sample size. It is true, that depending upon how these domains are defined, some areas of study may be too small for a reliable survey to be designed. However, other constraints could also be at play like budget limitations. 
Since design-based direct estimators fail to produce reliable estimates for small areas, model-based ``indirect'' estimators have become popular. These estimators borrow information from related areas, variables, and/or time periods thereby increasing the effective sample size for small area estimation problems. \cite{Datta2009}, \cite{Pfeffermann2013}, and \cite{Rao2015} provide comprehensive review of popular model-based estimators.

The Fay-Herriot model (FH; \citet{FH1979}) is one of the most popular small area models available at the area level. It was originally used to produce estimates of income in small areas with a population of less than 1000 and used auxiliary variables from different data sources such as administrative records and censuses. However, the original model formulation made the assumption that these auxiliary data were measured without any error. This assumption is not always valid and subsequent applications of such data have shown significant sampling error associated with them. 

To varying degrees, almost all statistical data contain measurement errors. The error may be a result of self-reporting, survey design, incorporating variables from other data sources, etc. The larger the sample, the smaller the level of sampling error, hence intuitively the smaller the geography (domain) we consider, the higher the impact of measurement error in the reported estimates. 


During model fitting often the assumption is made that the covariates do not contain any error and what we observe is the true value of the auxiliary variable. Failure to account for this error in the modeling framework can lead to biased, inconsistent estimates of regression parameters, severely impede inference, and often result in unreliable predictions \citep{Ogden2006, Caroll2006}.

A frequentist approach to the FH model on small areas with covariate measurement error was proposed by \cite{YL2008}. They applied their model to data from the 2003 to 2004 US National Health and Nutrition Examination Survey (NHANES), using the 2004 US National Health Interview Survey (NHIS) as auxiliary information, and showed that the use of the auxiliary information through the measurement error model resulted in substantial gain in estimated precision for small domains. In \cite{Arima2015}, the authors adopted a Bayesian implementation of this frequentist model with covariate measurement error and showed that a Bayesian adaptation of their model was significantly more stable in terms of uncertainty quantification in the estimates, and in \cite{Arima2017} they considered a multivariate FH model under a functional measurement error approach. 

Several recent authors have concentrated on measurement error in covariates in a variety of settings. \cite{Li2009} proposed a new class of linear mixed models for spatial data in the presence of covariate measurement errors and showed that, if measurement error is ignored the naive estimators of the regression coefficients are attenuated while the naive estimators of the variance components are inflated. \cite{Huque2014} showed that the parameter estimates could be highly sensitive to the choice of the spatial correlation structure when the covariates are measured with error. 
\cite{Matos2018} introduced a multivariate measurement error model when both the predictors and responses are censored.  \cite{Tadayon2018} introduced a general class of spatial models with covariate measurement error for non-Gaussian geostatistical data using a likelihood-based approach. 
\cite{Cabral2021} proposed measurement error models under the class of scale mixtures of skew-normal distributions to deal with case where the variables in the latent process depart from the normality assumption. Most of the existing literature on measurement error models involve a Gaussian response in the spatial or non-spatial settings. 
To our knowledge, estimating a non-Gaussian response using areal spatial covariates which are measured with error is lacking. 

Our goal in this paper is to extend the measurement error model paradigm to include both Gaussian and non-Gaussian spatial data at the area level. We propose a multivariate multi-type Bayesian hierarchical mixed effect model for area-level data which can be distributed from multiple classes of distributions both Gaussian or non-Gaussian, using auxiliary data which are measured with error and are spatially dependent, under Hierarchical Generalized Transformation \citep[HGT;][]{Bradley2021} model specification. Detailed model specifications will be discussed in subsequent sections, but for now, we shall motivate the novelty and application of our proposed model. 

The first aspect of our proposed model is that it arises from non-Gaussian data. The generalized linear mixed effects model (GLMM) is a standard approach to model non-Gaussian data. However, GLMMs often lead to non-conjugate distributions which can prove difficult to implement with high-dimensional datasets. A recent alternative is the latent conjugate multivariate (LCM) model \citep{BHW2018, Bradley2019}. A traditional approach here would be to impose a transformation on this data (e.g., a log transformation or a transformation in the Box–Cox family) so that it can be modeled using a mixed effect model framework or some other preferred model. The question remains on how to determine what is an appropriate transformation for our data. \cite{Bradley2021} proposed a Bayesian solution to determine this appropriate transformation using a hierarchical generalized transformation (HGT) model. Using the usual data models and appropriate transformation priors,
a non-Gaussian response is transformed into a dataset with continuous support so that it can be modeled using a preferred model like a latent Gaussian process (LGP) model. \cite{Bradley2021} introduced a joint Bayesian analysis of multi-type response datasets. By multi-type response we mean datasets with responses from multiple classes of distribution and using the HGT specification in our framework, we can simultaneously model them under a common latent process model. From our simulation studies and real data examples, we observe that in some cases joint modeling of multi-type responses helps improve the estimation process in terms of prediction error, compared to when they are modeled as single response data. These multi-type responses can be both Gaussian or non-Gaussian and the HGT component is flexible enough to handle both.  

Another aspect of our proposed model is the spatial covariates that have been measured with error. We assume that among the set of covariates we use, a subset or all of them have measurement error. For such covariates, we do not observe the true values, 
rather we observe an error-prone estimate. 
We assume the error occurs at random and model the observed estimate as a random variable whose mean is equal to the unobserved variable and variance equal to the measurement error variance in a \textit{classical} measurement error model \citep{Caroll2006}. 
While several spatial measurement error models have been proposed, a frequent assumption in them is a common error structure across the spatial domain \citep[see, e.g.][]{Huque2014, Tadayon2018}. It thus falls short of addressing the heteroscedasticity of spatio-temporal data. Another common assumption is that the underlying true covariate is smooth and whatever variability is observed is due to the measurement error. We have assumed that the error-prone covariates in our model are also spatially dependent and we impose a conditional autoregressive (CAR) model on them to account for the spatial variability in the estimation process. Furthermore, we assume that the measurement error variance itself is spatially varying. This is a departure from the existing common error covariance assumption. 
 
The rest of this article is organized as follows. In Section~\ref{Section: Methodology}, we discuss the modeling framework, 
and provide the necessary technical development behind the model hierarchy including details surrounding the model implementation. In Section~\ref{Section: Results} we present some analysis results from simulation studies designed to illustrate the model performance in different scenarios including both the single response and multiple response datasets. 
In Section~\ref{Section: RealEgs}, we illustrate potential applications of our proposed model under both single and multi-type response scenarios to U.S. Census Bureau's `Small Area Income and Poverty Estimates' (SAIPE) program and the U.S. Department of Housing and Urban Development's (HUD) housing assistance programs using county-level ACS data from four states in the U.S.  
Section~\ref{Section: Discussion} concludes this paper with a discussion of the analysis results we obtained in the previous sections and possible future directions. The derivation of full conditional distributions is provided in the Appendix. 
 
\section{Methodology} \label{Section: Methodology}
\subsection{The Hierarchical Generalized Transformation Model} \label{Section: HGT}
In this section, we discuss the first component of our model specification, which addresses the computational difficulty of modeling non-Gaussian data using existing Bayesian statistical tools. For specificity, we limit our initial discussion to areal count-valued survey responses, but later show that our specification can also include multi-type responses. 

Assume that $Z_i$ is a count-valued observation for the areal unit $i$. Then, if we assume conditional independence of $Z_i$ given an unobserved latent process $Y_i$, we have the following hierarchical structure
\begin{linenomath*}
\begin{equation}
    Z_i|Y_i\overset{ind.}{\sim} \hbox{Poisson}(\exp{\{Y_i\}}),~~~~~i=1,\dots,N. \label{GLMM}
\end{equation}
\end{linenomath*}
The Latent Gaussian Process (LGP) model is a standard tool for modeling such dependent count-valued data but their use leads to non-conjugate full-conditional distributions, and hence cannot be sampled from directly, thus often requiring
computationally expensive Metropolis-Hastings or other accept-reject algorithms. These 
modeling strategies often become difficult or computationally expensive to implement with modern high-dimensional datasets \citep{BHW2018, Bradley2020}.
Another traditional approach to modeling such non-Gaussian data would be to impose a
transformation such that 
\begin{linenomath*}
\begin{equation*}
    h(Z_i)|Y_i,\boldsymbol{\theta}\overset{ind.}{\sim} f(h(Z_i)|Y_i,\boldsymbol{\theta}),~~~~~i=1,\dots,N.
\end{equation*}
\end{linenomath*}
where $h(\cdot)$ is a transformation of the data $Z_i$ and $f(\cdot)$ is some ``preferred
model" as determined by the researchers. $h(Z_i)$ is conditionally independent of the latent
process $Y_i$ and the parameter set $\boldsymbol{\theta}$. \cite{Bradley2021} aims to solve the problem of determining this unknown transformation and developed a
Bayesian implementation of the Hierarchical Generalized Transformation (HGT) model. Their
implementation utilizes three main components: 1) ``Data Model": Distribution of the data given
a transformation, $f(Z_i|h_i)$, 2) ``Transformation Prior": Prior distribution of the
transformation, $f(h_i|\gamma)$, where $\gamma$ is a vector valued hyperparameter, 3)
``Transformed data model": The density $f(\textbf{h}|\textbf{y},\boldsymbol{\theta})=\prod_i
f(h(Z_i)|Y_i,\boldsymbol{\theta})$, and the ``Process Model":
$f(\textbf{y}|\boldsymbol{\theta})$, where $\textbf{h}=(h(Z_1),\ldots,h(Z_N))^{'}$. For simplicity of exposition, we write $h_i=h(Z_i)$ in all subsequent discussions. The hierarchical structure of this Bayesian implementation of the HGT model for a Poisson response takes the form 
\begin{linenomath*}
\begin{equation}
\begin{split}
    &Data\ Model:\  Z_i|h_i \overset{ind}{\sim} \hbox{Poisson}(\exp{\{h_i\}}), \ \ \ \ i=1,\ldots, N\\
    &Transformed\ Data\ Model:\ \mathbf{h}|\mathbf{y},\boldsymbol{\theta},\boldsymbol{\gamma} \propto f(\mathbf{h}|\mathbf{y},\boldsymbol{\theta})m(\mathbf{h}| \boldsymbol{\gamma}).
\end{split} \label{HGT}
\end{equation}
\end{linenomath*}
The $f(\cdot)$ in the transformed data model represents the ``preferred model" in HGT literature and will be discussed in the next section. The $m(\mathbf{h}| \boldsymbol{\gamma})$ in the transformed data model is a proportionality constant defined as 
\begin{equation*}
    m(\mathbf{h}|\boldsymbol{\gamma}) = f(\mathbf{h}|\boldsymbol{\gamma})\big /  \int \int f(\mathbf{h}|\mathbf{y},\boldsymbol{\theta})f(\mathbf{y}|\boldsymbol{\theta})f(\boldsymbol{\theta})d\mathbf{y} d \boldsymbol{\theta},
\end{equation*}
where $f(\mathbf{h}|\boldsymbol{\gamma})$ is the transformation prior that we will discuss later in the section. The presence of  $m(\mathbf{h}|\boldsymbol{\gamma})$ ensures that the marginal distribution of $\mathbf{h}$ is uniquely defined by our choice of transformation prior and the data model, and in fact neither changes the preferred model proportionally as a function of $\mathbf{y}$ and $\boldsymbol{\theta}$ nor has an effect on the MCMC updates of $\mathbf{y}$ or $\boldsymbol{\theta}$ \citep{Bradley2021}. The data model in \eqref{HGT} is different from that in \eqref{GLMM}. As opposed to conditioning on the latent process of interest in \eqref{GLMM}, in the HGT specification, we condition on the unknown transformation $h_i$. Thus, the hierarchy in \eqref{HGT} implies that the data $Z_i$ are conditionally independent given $h_i$. 

The data model in \eqref{HGT} can also be defined for continuous-valued data, $Z_i$ with variance $\nu$, as
\begin{equation}
    Data\ Model:\  Z_i|h_i,\nu \overset{ind}{\sim} \hbox{Normal}(h_i,\nu), \ \ \ \ i=1,\ldots, N \label{HGT2}
\end{equation}
or for integer-valued data ranging from $0,\ldots,b_i$ as 
\begin{equation}
    Data\ Model:\  Z_i|h_i \overset{ind}{\sim} \hbox{Binomial} \left \{ b_i,\frac{\exp{(h_i)}}{1+\exp{(h_i)}} \right \}, \ \ \ \ i=1,\ldots, N. \label{HGT3}
\end{equation}
The HGT model can be used to model each of the data models in \eqref{HGT},\eqref{HGT2}, and \eqref{HGT3} separately as single response type datasets, or can be jointly modeled as a multi-type response dataset, and in that case, the Bayesian implementation of the HGT model for the multi-type responses takes the form,
\begin{equation}
\begin{split}
 & Data\ Model\ 1:\ Z_{i1}|h_{i1}, \nu \overset{ind}{\sim} \hbox{Normal} (h_{i1},\textit{v})\\
 & Data\ Model\ 2:\ Z_{i2}|h_{i2} \overset{ind}{\sim} \hbox{Poisson} (\exp{\{h_{i2}\}})\\
 & Data\ Model\ 3:\ Z_{i3}|h_{i3} \overset{ind}{\sim} \hbox{Binomial}\left \{ b_i, \frac{\exp{\{h_{i3}\}}}{1+\exp{\{h_{i3}\}}} \right \}, \ \ i=1,\ldots, N, \ j=1,\ldots, 3\\
 & Transformed\ Data\ Model:\ \mathbf{h}|\mathbf{y},\boldsymbol{\theta},\boldsymbol{\gamma} \propto f(\mathbf{h}|\mathbf{y},\boldsymbol{\theta})m(\mathbf{h},\boldsymbol{\gamma}).
\end{split} \label{mHGT}
\end{equation}
 In \eqref{mHGT}, the transformed data vector $\mathbf{h}=(h_{11},\ldots,h_{N3})^{'}$ is of length $N^{*}(=j\times N)$ and the latent process $\mathbf{y}=(Y_{11},\ldots,Y_{N1},Y_{12},\ldots,Y_{N2},Y_{13},\ldots,Y_{N3})^{'}$ is $N^{*}$-dimensional.
 
 The transformation prior is defined to be a conjugate distribution associated with the data model. \cite{Bradley2021} shows that the conjugate distribution for $h_{ij}$ $(i=1,\ldots,N, j=1,\ldots,3)$ is of the form 
 \begin{equation*}
     f_{DY}(h_{ij}|\alpha_j,\kappa_j,a,b) = K(\alpha_j,\kappa_j)\exp{\{\alpha_jh_{ij}-\kappa_j\psi_j(h_{ij})\}},\ i=1,\ldots,N,\ j=1,\ldots,3
 \end{equation*}
where $K(\alpha_j,\kappa_j)$ is a normalizing constant, and $\alpha_1\in \mathbb{R}$, $\kappa_2>\alpha_2$, $\alpha_q>0$, and $\kappa_k>0$ for $q=2,3$, and $k=1,3$. For $\psi_1(Z)=Z^2$, $\psi_2 (Z)=\exp{(Z)}$, and $\psi_3(Z)=\log(1+e^Z)$, the transformation prior in the HGT model is $\hbox{DY}(\alpha_j,\kappa_j, \psi_j)$, $(j=1,2,3)$,  where the shorthand $\hbox{DY}(\cdot)$ represents the pdf associated with the \cite{Diaconis1979} prior. The set $\boldsymbol{\gamma}=\left [ \alpha_1,\alpha_2,\alpha_3,\kappa_1,\kappa_2,\kappa_3 \right ]^{'} $ represents the set of transformation hyperparameters. The data models in \eqref{mHGT} and the DY prior as above can be used to obtain the full conditional distribution for the elements of the transformations $\mathbf{h}$ as
\begin{equation}
\begin{split}
    h_{i1}|Z_{i1},\boldsymbol{\gamma} &\overset{ind.}{\sim} \hbox{Normal} \left \{ \left ( 2\kappa_1+\frac{1}{\textit{v}}\right )^{-1}\left ( \frac{Z_{i1}}{\textit{v}}+\alpha_1\right ), \left ( 2\kappa_1+\frac{1}{\textit{v}}\right )^{-1} \right \}, \\
    h_{i2}|Z_{i2},\boldsymbol{\gamma} &\overset{ind.}{\sim} \hbox{DY}(\alpha_3+Z_{i2},\kappa_2+1;\psi_2) , \\
    h_{i3}|Z_{i3},\boldsymbol{\gamma} &\overset{ind.}{\sim} \hbox{DY}(\alpha_2+Z_{i3},\kappa_2+b_i;\psi_3);~~~~~ i=1,\ldots, N, \ j=1,\ldots, 3.
\end{split} \label{eq HGT}
\end{equation}
Posterior replicates of $h_{ij}\ $ $(j=1,2,3)$ can be computed using
\begin{equation*}
\begin{split}
    h_{i1} &= \left ( 2\kappa_1+\frac{1}{\textit{v}}\right )^{-1}\left ( \frac{Z_{i1}}{\textit{v}}+\alpha_1\right ) + \omega_1,  \\
    h_{i2} &= \log (\omega_2),\\
    h_{i3} &= \log\left ( \frac{\omega_2}{1-\omega_3}\right );~~~~~ i=1,\ldots, N, \ j=1,\ldots, 3
\end{split}
\end{equation*}
where 
$\omega_1| Z_{i1},\alpha_1,\kappa_1,\textit{v}\sim \hbox{Normal}\left ( 0, \left ( 2\kappa_1+\frac{1}{\textit{v}}\right )^{-1} \right )$, 
$\omega_2|Z_{i2},\alpha_2,\kappa_2 \sim \hbox{Gamma}(\alpha_2+Z_{i2},\kappa_2+1)$, and $\omega_3|Z_{i3},\alpha_3,\kappa_3 \sim \hbox{Beta}(\alpha_3+Z_{i3},\kappa_3-\alpha_3+b_i-Z_{i3})$.
See \cite{Bradley2020} and \cite{Bradley2021} for further details about the $\hbox{DY}(\cdot)$ transformation priors and the HGT model specifications. 

The motivation behind utilizing this HGT model 
is that now the transformed dataset can be used in place of the original multi-type response dataset(s) while implementing the ``preferred model", without facing the computational difficulties in a MCMC sampler. The HGT component of our model 
also gives us a framework to either model a single type response or jointly model multi-type response datasets under a unified model specification using the `preferred' model defined in Section~\ref{Section: HGT-SME}.


\subsection{The HGT-Spatial Mixed Effect Measurement Error Model} \label{Section: HGT-SME}
Utilizing the HGT component of our model specification as described in the previous section, we have transformed the Gaussian or non-Gaussian response(s), into a continuous response data $\mathbf{h}$. 
We define the following mixed effect model for the transformed data
\begin{equation}
    \textbf{h}|\boldsymbol{\beta},\boldsymbol{\delta,\eta,\xi,\gamma},\textbf{W,M},\tau^2 \propto \hbox{Gaussian}(\textbf{W}\boldsymbol{\beta}+\mathbf{S}\boldsymbol{\delta}+\textbf{M}\boldsymbol{\eta}+\boldsymbol{\xi},\tau^2\mathbf{I}) m(\textbf{h}|\boldsymbol{\gamma}).
\end{equation}
Here, $\mathbf{W}=(\mathbf{w}_1^{(j)},\ldots,\mathbf{w}_{N}^{(j)})^{'}$ is a $p$-dimensional design matrix consisting of the $p\times 1$ vector of unobserved covariates $\textbf{w}_i^{(j)}=(w_{i,1}^{(j)},\ldots,w_{i,p}^{(j)})^{'}$ that are measured with error for each response type $j$, and $\mathbf{S}=(\mathbf{s}_1^{(j)},\ldots,\mathbf{s}_N^{(j)})^{'}$ is a $q$-dimensional matrix of $q\times 1$ vector of covariates $\mathbf{s}_i^{(j)}=(s_{i,1}^{(j)},\ldots,s_{i,q}^{(j)})^{'}$ that are measured without error for each response type $j$. $\mathbf{S}$ is assumed independent of $\mathbf{W}$. If an intercept is required in the model, it suffices to include $1$ in $\mathbf{s}_i^{(j)}$. The vector-valued parameters $\boldsymbol{\beta} \in \mathbb{R}^p$ and $\boldsymbol{\delta} \in \mathbb{R}^q$ are the unknown coefficient parameters specified to have a Normal prior with mean $0$, and variance parameter $\sigma^2_{\beta}$ and $\sigma^2_{\delta}$ respectively. 

The multivariate spatial dependencies are accounted for by the true covariates $\mathbf{W}$, and therefore we design the $\mathbf{M}\boldsymbol{\eta}$ component to account for any residual spatial dependency. The $r$-dimensional random coefficient vector $\boldsymbol{\eta}$, is assumed to follow a Gaussian distribution with mean zero, and unknown variance  $\sigma^2_{\eta}$.  The matrix $\mathbf{M}$ consists of $r-$dimensional vectors of multivariate spatial basis functions $\mathbf{M}_i^{(j)}=(M_1^{(j)},\ldots,M_r^{(j)})^{'}$ for $i=1,\ldots, N,\ j=1,2,\ldots$, with $r\ll N$. The vector $\boldsymbol{\xi}=(\xi_1^{(j)},\ldots,\xi_N^{(j)})^{'}$ represents the fine-scale variation and are assumed to be independent Gaussian random variables with mean zero and variance  $\sigma^2_{\xi}\mathbf{I}$. In general, this represents the leftover variability not accounted for by random effect components $\textbf{W}\boldsymbol{\beta}+\mathbf{M}\boldsymbol{\eta}$ and the fixed effect components $\mathbf{S}\boldsymbol{\delta}$. 

We do not observe $\mathbf{W}$, but rather an estimate $\mathbf{X}=(\mathbf{x}_1^{(j)},\ldots,\mathbf{x}_N^{(j)})^{'}$ where $\textbf{x}_i^{(j)}=(x_{i,1}^{(j)},\ldots,x_{i,p}^{(j)})^{'}$. We assume this estimate is prone to additive measurement error and, thus, adopt a classical measurement error model \citep{Caroll2006, Fuller2009}, as 
\begin{linenomath*}
\begin{equation}
   x_{i,k}^{(j)}=w_{i,k}^{(j)}+u_{i,k}^{(j)}, \ \ \ i=1,\ldots,N,\ k=1,\ldots,p,\ j=1,2,\ldots
\end{equation}
\end{linenomath*}
where $\mathbf{U}=(\mathbf{u}_1^{(j)},\ldots,\mathbf{u}_N^{(j)})^{'}$ is the measurement error component and for each error prone covariate $\mathbf{w}_k^{(j)} = (w_{1,k}^{(j)},\ldots,w_{N,k}^{(j)})^{'}\ (k=1,\ldots,p)$, we assume the measurement error $\mathbf{u}_k^{(j)} = (u_{1,k}^{(j)},\ldots,u_{N,k}^{(j)})^{'}\ (k=1,\ldots,p)$ to have mean $\mathbf{0}$ and covariance matrix $\boldsymbol{\Sigma}(\mathbf{u}_k^{(j)})$. Here we assume $\boldsymbol{\Sigma}(\mathbf{u}_k^{(j)})$ to be known and fixed, but it can be treated as an unknown parameter for estimation  \citep[see, e.g.,][]{Caroll2006}. 
Our choice of a \textit{classical} measurement error structure over a Berkson error is guided by our motivating data examples involving survey tabulations. In survey data like that from ACS, we assume that the measurement error occurs independent of the true value and that the observed measurements are the true values with additive measurement error. We assume that the true value is not fixed and instead varies across space. 

In order to account for the spatial dependence in the covariate measured with error, $\mathbf{W}$, we have assumed a Gaussian prior on $\mathbf{w}_k^{(j)}$ 
with mean $\boldsymbol{\mu}({\mathbf{w}_{k}^{(j)}})$ and covariance matrix $\boldsymbol{\Sigma}({\mathbf{w}_k^{(j)}})=\sigma^2_{\mathbf{w}_k}(\mathbf{D}-\rho_k \mathbf{A})^{-1}$ with $\mathbf{A}$ being the adjacency matrix (i.e., $a_{ll}=0$ and $a_{lm}=1$ if $l\sim m$ or $0$ otherwise, where the notation $l\sim m$ refers to the case when $l$ is a neighbor of $m$ and $m$ is a neighbor of $l$) and $\mathbf{D}$ a diagonal matrix with entries equal to the number of neighbors for location $l$. This represents a Conditional Autoregressive (CAR) prior on $\mathbf{w}_k^{(j)}$, $(k=1,\ldots,p)$ \citep[see, e.g.,][] {Besag1974, Cressie2011, BanerjeeGelfand2015}. The mean $\boldsymbol{\mu}({\mathbf{w}_{k}^{(j)}})$ could be set to zero, or it can be estimated via a regression model itself. 
$\rho_k$ is the spatial autocorrelation parameter associated with $\mathbf{w}_k^{(j)} \ (k=1,\ldots, p)$, and in order to guarantee the non-singularity of $(\mathbf{D}-\rho_k \mathbf{A})^{-1}$, we assume $|\rho_k|<1$. 
A Uniform$(a,b)$ prior can be specified on $\rho_k$, with limits $a$ and $b$ to be determined depending on the context. 
For $p>1$, we have two possible scenarios. If $\mathbf{w}_k^{(j)}$ $(k=1,\ldots,p)$ are assumed mutually independent of each other, then, their prior distributions would be the univariate CAR model as defined above and as it appears in the Process Model 1 in the model hierarchy in \eqref{Eq. HGT-SME}. If they are assumed dependent, then a multivariate CAR (MCAR) model is defined. \cite{Gelfand2003}, \cite{Jin2005,Jin2007}, \cite{Porter2014a}, and \cite{BanerjeeGelfand2015} provide popular specifications of MCAR model.

 The proposed modeling framework is flexible enough to incorporate different types of basis functions  \citep[see,][]{Bradley2011}. We use the Moran's I (MI) basis
functions \citep{Grif2000}. Our motivation in using these basis functions come primarily from their properties related to confounding in spatial mixed effect models \citep{Reich2006}, as well as it's use for dimension reduction \citep{HH2013, BHW2015} since the proposed framework can be calibrated to big data models.
The MI operator is explicitly defined as
\begin{linenomath*}
\begin{equation}
    G(\mathbf{L},\mathbf{A})\equiv (\mathbf{I}_{N^{*}}-\mathbf{L}(\mathbf{L}^{'}\mathbf{L})^{-1}\mathbf{L}^{'})\mathbf{A}(\mathbf{I}_{N^{*}}-\mathbf{L}(\mathbf{L}^{'}\mathbf{L})^{-1}\mathbf{L}^{'}) \label{Eq: MIoperator}
\end{equation}
\end{linenomath*}
where $\mathbf{L}=\left [ \mathbf{W}\ \mathbf{S}\right ]$, 
$\mathbf{I}_{N^{*}}$ is an $N^{*}\times N^{*}$ identity matrix, and $\mathbf{A}$ is the adjacency matrix for the given spatial geography. The MI operator in \eqref{Eq: MIoperator} defines a column space that is orthogonal to $\mathbf{L}$ and thus guarantees there are no confounding issues between the random and fixed effect coefficients.  $\mathbf{M}$ is comprised of the first `$r$' columns of $\Phi_{\mathbf{L},G}$ which is obtained from the spectral representation $G(\mathbf{L},\mathbf{A})=\Phi_{\mathbf{L},G} \Lambda_{\mathbf{L},G}\Phi^{'}_{\mathbf{L},G}$  (see \citet{BHW2015} for more details). This allows for faster computation of the distribution for the random effects $\boldsymbol{\eta}$ in a reduced dimensional space.  We note here that although we motivate our choice of MI basis function, in part, for its dimension reduction properties other choice of basis functions like Obled-Creutin basis functions \citep{Obled1986, Bradley2017} could also be used and are optimal from a dimension reduction perspective. 

Prior distributions for the variance components $\sigma^2_{\mathbf{w}_k}$, $\sigma^2_{\eta}$, and $\sigma^2_{\xi}$ are specified so that the conjugacy, with the transformed data model and the process models is maintained and we can obtain exact expressions for the full conditional distributions within a Gibbs sampling algorithm. In particular, we chose inverse-gamma $IG (\alpha,\beta)$ priors. $\boldsymbol{\tau^2}$ represents the sampling error variance of the response on the transformed space. In most cases the statistical agencies will provide estimates of the sampling errors and they can be used to inform the choice of prior distributions. 
For our model, since the HGT model handles the unknown transformation for the multi-type data, it can also be used to transform the sampling error following appropriate changes in the transformation priors, but care has to be taken regarding interpretation of the sampling error in a joint modeling of multi-type responses. 
This 
is left as a subject of future research. 
We treat $\boldsymbol{\tau^2}$ as a model parameter to be estimated and assume an $IG (\alpha,\beta)$ prior.

The latent Gaussian spatial mixed effect measurement error model we have considered, which henceforth will be referred to as the HGT-SME model, has the following hierarchical structure: 
\begin{eqnarray}
\begin{split}
       &Transformed\ Data\ Model:\\ 
       &\quad \mathbf{h}|\mathbf{W},\boldsymbol{\beta,\delta},\boldsymbol{\eta,\gamma,\xi},\textbf{M},\tau^2 \propto \hbox{Gaussian}(\textbf{W}\boldsymbol{\beta}+\mathbf{S}\boldsymbol{\delta}+\textbf{M}\boldsymbol{\eta}+\boldsymbol{\xi},\tau^2\mathbf{I}) \times m(\mathbf{h}|\boldsymbol{\gamma}),\\
        & Date\ Model\ 4: \ \mathbf{X|W,\Sigma_U} \sim \hbox{Gaussian}(\mathbf{W},\Sigma_U)\\
        & Process\ Model\ 1: \ \mathbf{W}|\boldsymbol{\mu}_W,\sigma^2_W,\boldsymbol{\Sigma}_W\sim \hbox{Gaussian}(\boldsymbol{\mu}_W,\sigma^2_W\boldsymbol{\Sigma}_W)\\
        & Process\ Model\ 2: \ \boldsymbol{\eta|\sigma^2_{\eta}}\overset{ind.}{\sim} \hbox{Gaussian}(\mathbf{0}_r,\sigma^2_{\eta}\mathbf{I}_r)\\
        & Process\ Model\ 3: \ \boldsymbol{\xi|\sigma^2_{\xi}}\overset{ind.}{\sim} \hbox{Gaussian}(\mathbf{0}_N,\sigma^2_{\xi}\mathbf{I}_N)\\
        & Prior\ 1: \ \boldsymbol{\beta} \sim \hbox{Gaussian}(\mathbf{0}_p,\sigma^2_{\beta}\mathbf{I}_p)\\
        & Prior\ 2: \ \boldsymbol{\delta} \sim \hbox{Gaussian}(\mathbf{0}_q,\sigma^2_{\delta}\mathbf{I}_q)\\
        & Prior\ 3: \ \boldsymbol{\tau^2} \sim \hbox{IG}(\alpha_{\tau},\beta_{\tau})\\
        & Prior\ 4: \ \boldsymbol{\sigma^2_{\xi}} \sim \hbox{IG}(\alpha_{\xi},\beta_{\xi})\\
        & Prior\ 5: \ \boldsymbol{\sigma^2_{\eta}} \sim \hbox{IG}(\alpha_{\eta},\beta_{\eta})\\
        & Prior\ 6: \ \boldsymbol{\sigma^2_{W}} \sim \hbox{IG}(\alpha_{W},\beta_{W}). 
\end{split} \label{Eq. HGT-SME}
\end{eqnarray}

\subsection{Model Implementation} \label{Implementation}
The HGT-SME model framework is general enough so that it can be applied to both single or multi-type responses, single or multiple covariates both with and without measurement error, spatially or spatio-temporally dependent or independent, and/or any other structure on the covariate space. Our choice of basis functions is meant for illustration and, in part, is based on their dimension reduction capability given that we are applying our proposed models to estimate survey responses which are often high dimensional. 
The framework is flexible enough to handle different basis functions and prior distributions that may prove to be advantageous in the context of different problems.

Assuming the model hierarchy in 
\eqref{Eq. HGT-SME}, we can derive full conditional distributions all of which have a closed form and are discussed in the Appendix. The collapsed Gibbs sampler simply samples cyclically from $f(\textbf{h}|\boldsymbol{\gamma}, \textbf{z})$, $f(\boldsymbol{\gamma}|\textbf{h}, \textbf{z})$, and the full conditional distributions of $\textbf{W}, \boldsymbol{\beta}, \boldsymbol{\delta}, \boldsymbol{\eta}, \boldsymbol{\xi}, \sigma_{w}^{2}, \sigma_{\xi}^{2}, \sigma_{\eta}^{2}$, and $\tau^{2}$ at every iteration. The updates for $\textbf{h}$ and $\boldsymbol{\gamma}$ collapse across the other processes and parameters and can be sampled from directly according to \eqref{eq HGT}. 
The sampler draws 10,000 replicates for each parameters from their full conditional distributions, out of which 5000 replicates are discarded as burn-in. Convergence of the MCMC algorithm was assessed visually using trace plots of the sample chains for each parameter, with no lack of convergence detected. 
We found that the results are robust to the specifications of the values for the hyperparameters and we have set $\sigma^2_{\beta}=\sigma^2_{\delta}=100$, $\alpha_{\tau},\alpha_{\eta},\alpha_{\xi},\alpha_{W}=2$, and $\beta_{\tau},\beta_{\eta},\beta_{\xi},\beta_{W}=1$ so as to ensure relatively noninformative (vague) priors on the scale of the data. We set $\boldsymbol{\mu}({\mathbf{w}_{k}^{(j)}})=0$. Setting $\rho=1$ in $\Sigma({\mathbf{w}_{k}^{(j)}})=(\mathbf{D}-\rho_k \mathbf{A})^{-1}$ leads to computational issues owing to  $\Sigma_{W}$ being singular. As such, \cite{BanerjeeGelfand2015} suggest setting $\rho \in (1/\lambda_{(1)},1/\lambda_{(n)})$, where $\lambda_{(1)}< \cdots < \lambda_{(n)}$ are the ordered eigenvalues of $\mathbf{D}^{-1/2}\mathbf{A}\mathbf{D}^{-1/2}$ to make $\Sigma_{W}$ non-singular. 
\cite{Wall2004} and \cite{BanerjeeGelfand2015} however show that setting $\rho$ closer to 1 leads to a more moderate association measure like Moran's I or Geary's C which are often used as an indicator for spatial dependency. 
In all subsequent sections, we shall refer to our proposed model as the HGT-SME and the case where measurement error has been ignored in the covariates as the \textit{naive} model. Thus, the \textit{naive} model has the following hierarchical structure
\begin{eqnarray}
\begin{split}
        &Transformed\ Data\ Model:\ \mathbf{h}|\boldsymbol{\beta},\boldsymbol{\eta,\gamma,\xi},\textbf{M},\tau^2 \propto \hbox{Gaussian}(\textbf{L}\boldsymbol{\beta}+\textbf{M}\boldsymbol{\eta}+\boldsymbol{\xi},\tau^2\mathbf{I}) m(\mathbf{h}|\boldsymbol{\gamma}),\\
        & Process\ Model\ 1: \ \boldsymbol{\eta|\sigma^2_{\eta}}\overset{ind.}{\sim} \hbox{Gaussian}(\mathbf{0}_r,\sigma^2_{\eta}\mathbf{I}_r)\\
        & Process\ Model\ 2: \ \boldsymbol{\xi|\sigma^2_{\xi}}\overset{ind.}{\sim} \hbox{Gaussian}(\mathbf{0}_N,\sigma^2_{\xi}\mathbf{I}_N)\\
        & Prior\ 1: \ \boldsymbol{\beta} \sim \hbox{Gaussian}(\mathbf{0}_p,\sigma^2_{\beta}\mathbf{I}_p)\\
        & Prior\ 2: \ \boldsymbol{\tau^2} \sim \hbox{IG}(\alpha_{\tau},\beta_{\tau})\\
        & Prior\ 3: \ \boldsymbol{\sigma^2_{\xi}} \sim \hbox{IG}(\alpha_{\xi},\beta_{\xi})\\
        & Prior\ 4: \ \boldsymbol{\sigma^2_{\eta}} \sim \hbox{IG}(\alpha_{\eta},\beta_{\eta}).
\end{split} \label{Eq. naive}
\end{eqnarray}
where $\mathbf{L}=[\mathbf{X}, \mathbf{S}]$ is the $N^{*}\times (p+q)$ design matrix consisting of covariates which are assumed to have no measurement error, with $\mathbf{X}$ and $\mathbf{S}$ as defined in Section~\ref{Section: HGT-SME}. $\boldsymbol{\beta}\in \mathbb{R}^{p+q}$ are the unknown coefficient parameters. 
All other parameters and model components have the same definition as discussed earlier in Section~\ref{Section: HGT-SME}. 
We will use this \textit{naive} model to compare the performance of our proposed model in subsequent sections.

\section{Empirical Simulation Studies} \label{Section: Results}
\subsection{The American Community Survey (ACS) Data}

The US Census Bureau conducts the American Community Survey (ACS), which is a national survey designed to provide reliable annual socio-economic, demographic, and housing data. With a sample size of approximately 3.5 million addresses, the ACS provides estimates over a 1-year period and 5-year period. ACS 1-year period estimates are available for geographic areas with at least 65,000 people. The Census Bureau combines 5 consecutive years of ACS data to produce multiyear period estimates for geographic areas with fewer than 65,000 residents \citep{Census2020}. 
Unlike the long form decennial census which is based on the entire population, ACS estimates are based on a sample and hence are subject to sampling error. To assess the impact of sampling error on the reliability of the data, the Census Bureau provides the ``margin of error” (MOE) for each published ACS estimate. 

Our model specification is motivated by the presence of measurement error in survey data. As such, we illustrate the application of our model specification with empirical simulation studies based on ACS estimates at the county level. For a single response scenario we concentrate on a count-valued response, and for the multiple response scenario we illustrate our model performance using a Poisson count valued and a Gaussian continuous valued response. In the following sections we describe the study design, 
followed by the analysis results. 

\subsection{Single-type Response Study}
We first evaluate the proposed HGT-SME model using a single response type data. We designed an empirical simulation study based on a ``pseudo" dataset which is generated based on the ACS 2019 5-year estimates for all counties in the states of Washington, Oregon, Idaho, and Montana. There are a total of $175$ counties in this study area. Specifically, let $Z_l$ denote the ACS 2019 5-year period (direct) estimate of the count of population below the poverty threshold for county $`l$' $(l=1,\ldots,N)$ in the study area, then a pseudo data value was generated as
\begin{linenomath*}
\begin{equation}
    Z_l^{*} \sim \hbox{Poisson}(Z_l +1), \ \ \ \ l=1,\ldots,N. \label{Pseudogen}
\end{equation} 
\end{linenomath*}
The data generated using \eqref{Pseudogen} is such that the Poisson random variable has as its mean the count-valued ACS direct estimates so that it very closely resembles our target dataset. The constant has been added to ensure that the mean of the Poisson random variables are also greater than 0. Thus, the $\mathbf{Z^{*}}$ is our count valued input into the HGT component of our model specification.

For this illustration, we designed the study where we have only one covariate that is measured with error, and set $p=1$. As our error-prone covariate, we use the log of the ACS 2019 5-year period estimates of the median household income per county in the study area. This covariate was spatially correlated as suggested by Moran's I of $0.39$ ($p$-value$ < 2.2e-16$). The log transformation was used to ensure that certain model assumptions are satisfied (e.g., Gaussianity, etc.). We use the corresponding standard error reported by ACS as the sampling error variance. The delta method was used to accommodate the log transformation of the covariate.  As the fixed effect covariate, we use the ACS 2019 5-year period estimates of the percentage of county population that received SNAP (Supplemental Nutrition Assistance Program). The model also includes an intercept, and thus we have $q=2$. We note here that although, this data is obtained from ACS and is prone to measurement error, only for the purpose of illustration we have assumed it to be free of measurement error. All other parameters and hyperparameters follow the model specifications detailed previously. The small area estimates are produced based on the posterior mean of the 5000 post burn-in samples from the Gibbs sampler. Given an estimate $\widehat{Z}^{*}_l$ for an areal unit $`l$', the root mean squared error (RMSE), and the mean absolute bias (also called mean absolute error or MAE), defined as 
\begin{equation*}
    RMSE = \sqrt{\frac{1}{N}\sum\limits_{l=1}^N (Z^{*}_l-\widehat{Z}^{*}_l)^2},\ \ \ \hbox{MAE}=\frac{1}{N} \sum\limits_{l=1}^N |Z^{*}_l-\widehat{Z}^{*}_l|
\end{equation*}
were computed to compare the improvement in predictions that are achieved when measurement error is considered in the model specification over when it is ignored. $50$ independent replicates of $\mathbf{Z}^{*}$ are simulated using \eqref{Pseudogen} for the study. 

\paragraph*{Choosing $\rho$ and $r$:} Following the suggestions in literature and based on some 
exploratory analysis (results not included here), 
we fixed $\rho=0.99$ instead of treating it as a parameter to be estimated for this study.  We adopt Moran's I basis functions. The rule of thumb suggests setting $r$ equal to approximately the top $10\%$ of the available basis functions \citep{HH2013}. However, initial analysis suggested the application requires a different specification and we chose three possible values of $r$ to be $67$, $35$, and $18$ for further evaluation, which correspond to $95\%$, $50\%$, and $25\%$ of the positive eigenvalues of the Moran's I operator. Here $N=175$ and, thus, the dimension of $\mathbf{M}$ is $175\times r$. 

\paragraph*{Study Results}
Table~\ref{table: ESS1.1} shows the median RMSE, MSE, and MAE values across all $50$ samples used in the simulation study.  We tabulate the results for all three values of $r$ and note that we get the lowest mean squared prediction error for $r=67$ which corresponds to $95\%$ of the eigenvalues. At this value of $r$, when we account for the measurement error in the covariate, we see an approximate $30\%$ reduction in the MSE and a corresponding $25\%$ reduction in MAE in the model estimates for the counties in the study area as compared to the naive model. This suggests that including the measurement error component in the model specifications improves the small area estimation in terms lower RMSE and MAE. 
\begin{table}[t] 
\renewcommand{\arraystretch}{1.2}
\setlength{\tabcolsep}{10pt}
\centering
\caption{Analysis results for Simulation Study 1.1. Values inside parenthesis represents relative $\%$ reduction in corresponding metric using HGT-SME model over the naive model. \label{table: ESS1.1}}
\resizebox{\columnwidth}{!}{%
\begin{tabular}{c|c|c|l|l|l}
\toprule
 \multicolumn{1}{ c| }{$\rho$} & \multicolumn{1}{ c| }{$r$} & \multicolumn{1}{ c| }{Estimator} & \multicolumn{1}{ c| }{RMSE} & \multicolumn{1}{ c| }{MSE} & \multicolumn{1}{ c }{Abs. Bias}\\ 
 \cline{1-6}
 \multicolumn{1}{ l|  }{\multirow{2}{*}{ }} & \multicolumn{1}{ l  }{\multirow{2}{*}{$67 \  (95\%)$ }}  & \multicolumn{1}{ |l| }{Naive model} & $5080.016$ & $2.58\times 10^{7}$ & $1654.958$\\ 
\multicolumn{1}{ c|  }{}& \multicolumn{1}{ c  }{} & \multicolumn{1}{ |l| }{HGT-SME} & $4260.933$ (\textcolor{red}{$\downarrow 16.1\%$}) & $1.82\times 10^7 $ (\textcolor{red}{$\downarrow 29.6\%$}) & $1243.958$ (\textcolor{red}{$\downarrow 24.8\%$}) \\ 
\cline{2-6}
\multicolumn{1}{ l|  }{\multirow{2}{*}{$0.99$ }} & \multicolumn{1}{ l  }{\multirow{2}{*}{$35\ (50\%)$ }}  & \multicolumn{1}{ |l| }{Naive model} & $5131.889$ & $2.63\times 10^7$ & $1700.634$ \\ 
\multicolumn{1}{ c|  }{}& \multicolumn{1}{ c  }{} & \multicolumn{1}{ |l| }{HGT-SME} & $4387.348$ (\textcolor{red}{$\downarrow 14.5\%$}) & $1.92\times 10^7$ (\textcolor{red}{$\downarrow 26.9\%$}) & $1280.891$ (\textcolor{red}{$\downarrow 24.7\%$}) \\ 
\cline{2-6}
\multicolumn{1}{ l|  }{\multirow{2}{*}{}} & \multicolumn{1}{ l  }{\multirow{2}{*}{$18 \ (25\%)$ }}  & \multicolumn{1}{ |l| }{Naive model} & $5421.715$  & $2.94\times 10^7$  & $1762.881$ \\ 
\multicolumn{1}{ c|  }{}& \multicolumn{1}{ c  }{} & \multicolumn{1}{ |l| }{HGT-SME}  & $4650.132$ (\textcolor{red}{$\downarrow 14.2\%$}) & $2.16\times10^7$ (\textcolor{red}{$\downarrow 26.4\%$}) & $1341.174$ (\textcolor{red}{$\downarrow 23.9\%$}) \\ 
 \cline{1-6} \end{tabular}%
 }
\end{table}

\subsection{Multi-type Response Study} \label{Section: SimStudy3}
We now evaluate the proposed HGT-SME model for a multi-type response dataset using a similar empirical simulation study. For this study, we limit ourselves to the Gaussian and Poisson response types. Hence, we have, $Z_{l1}$ as the continuous Gaussian response and $Z_{l2}$ as the count-valued Poisson response. A pseudo dataset was generated based on the ACS 2019 5-year county estimates in the states of Washington, Oregon, Idaho, and Montana. There are a total of $175$ counties in this study area. Specifically, if $Z_{l1}$ is the 2019 ACS 5-year period estimates of median housing cost for county $`l$' $(l=1,\ldots,N)$ in the study area, then a continuous pseudo data was generated as 
\begin{linenomath*}
\begin{equation}
    Z_{l1}^{*} \sim \hbox{Normal} (\log(Z_{l1}),1), \ \ \ \ l=1,\ldots,N. \label{Pseudogen2.1}
\end{equation} 
\end{linenomath*}
The log transformation was used to satisfy model assumptions like Gaussianity. If $Z_{l2}$ is the ACS 2019 5-year period estimate of the count of population below poverty threshold for county $`l$' $(l=1,\ldots,N)$ in the study area, then a count-valued Poisson pseudo data was generated as
\begin{linenomath*}
\begin{equation}
    Z_{l2}^{*} \sim \hbox{Poisson} (Z_{l2} +1), \ \ \ \ l=1,\ldots,N. \label{Pseudogen2.2}
\end{equation} 
\end{linenomath*}
The data generated using \eqref{Pseudogen2.1} and \eqref{Pseudogen2.2} acts as our input into the HGT component of the model hierarchy. Hence, $\mathbf{h}$ is a vector of length $N^{*}=2N$ comprising the transformed data with continuous support. Here too we assume a single covariate measured with error and one fixed covariate without any measurement error. The model also has an intercept and thus we have $p=1$ and $q=2$. As the error-prone covariate, we use the log of the ACS 2019 5-year period estimates of median household income per county in the study area and as the fixed covariate without any measurement error, we use the ACS 2019 5-year period estimates of the percentage of county population that received SNAP. 
All other parameters and hyperparameters follow the model specifications detailed earlier. The small area estimates are produced based on the posterior mean of the 5000 post burn-in samples from the Gibbs sampler. Given an estimate $\widehat{Z}^{*}_l$ for an areal unit $`l$', root mean squared error (RMSE) and the mean absolute bias (also called mean absolute error or MAE), were computed to compare the improvement in predictions that is achieved when measurement error is considered in the model specification over when it is ignored. $50$ independent replicates of $\mathbf{Z}^{*}_j$ $(j=1,2)$ are simulated using \eqref{Pseudogen2.1} and \eqref{Pseudogen2.2} for the study. 

\paragraph*{Choosing $\rho$ and $r$:} Following our observations from the simulation studies for the single response case, we fixed $\rho=0.99$ instead of treating it as a parameter to be estimated for these simulation studies.  We chose three possible values of $r$ to correspond to $95\%$, $50\%$, and $25\%$ of the eigenvalues of the Moran's I operator. Based on our observation from the single response case where the lowest RMSE and absolute bias were observed for $r$ corresponding to $95\%$ of the eigenvalues, we limit our discussion of the results with $r$ set to $95\%$ of the eigenvalues of the Moran's I operator. Here $N=175$ and thus dimension of $\mathbf{M}$ is $N^{*}\times r$. 

\paragraph*{Study Results:}
Table~\ref{table: ESS2.1} shows the median RMSE, MSE, and MAE values across all $50$ samples used in the simulation study. When we account for the measurement error in the covariate, for the Gaussian response, we observe an approximate $44\%$ reduction in MSE and a corresponding $25\%$ reduction in absolute bias in the model estimates as compared to the naive case. 
For the Poisson response, we observe an approximate $39\%$ reduction in MSE and a corresponding $32\%$ reduction in absolute bias for the case where we account for measurement error in the covariate. Similar trends were observed for lower values of $r$ but those results have been omitted as the case presented here sufficiently encapsulates the purpose of this study. Similar to the single response studies, we observe that including the measurement error specification in the model helps improve the small area estimation, and in fact jointly multi-type responses here, further improves the results over the single response study. 
\begin{table}[t]
\centering
\caption{Analysis results for Simulation Study 2.1.  Values inside parenthesis represents relative $\%$ reduction in corresponding metric using HGT-SME model over the naive model.\label{table: ESS2.1}}
\resizebox{\columnwidth}{!}{%
\setlength{\tabcolsep}{2pt}
\renewcommand{\arraystretch}{1.4}
\begin{tabular}{llll|lll}
\toprule
& \multicolumn{3}{ c| }{Gaussian Response} & \multicolumn{3}{ c }{Poisson Response}\\ \cline{1-7}
\multicolumn{1}{ c }{Estimator} & \multicolumn{1}{ c }{RMSE} & \multicolumn{1}{ c }{MSE} & \multicolumn{1}{ c| }{Abs. Bias} & \multicolumn{1}{ c }{RMSE} & \multicolumn{1}{ c }{MSE} & \multicolumn{1}{ c }{Abs. Bias} \\ \cline{1-7}
\multicolumn{1}{l|}{Naive model} & 0.4357 & 0.1898 & $0.3465$ & 4997.59 & $2.50 \times 10^7$ & 1636.978\\
\multicolumn{1}{l|}{HGT-SME} & 0.3262 (\textcolor{red}{$\downarrow 25\%$}) & 0.1064 (\textcolor{red}{$\downarrow 44\%$}) & $0.2583$ (\textcolor{red}{$\downarrow 25\%$}) & 3895.155 (\textcolor{red}{$\downarrow 22\%$}) & $1.52 \times 10^7$ (\textcolor{red}{$\downarrow 39\%$})  & 1115.470 (\textcolor{red}{$\downarrow 32\%$}) \\
\cline{1-7}
\end{tabular}%
}

\end{table}
\section{American Community Survey Applications}  \label{Section: RealEgs}
\subsection{The SAIPE Motivated Illustration} \label{Section: RealEg1}
In this section we apply our model to illustrate a real data application motivated by the U.S. Census Bureau's Small Area Income and Poverty Estimates (SAIPE) program. SAIPE is conducted annually to provide estimates of income and poverty statistics for the population in several age groups at the state, county and school district levels. Federal and local governmental agencies utilize these estimates to allocate federal funds, and manage development projects. In addition to this, the SAIPE also provides single-year poverty estimates for school aged population between ages 5-17 across all school districts in the U.S. 
Further information about the program and the data can be found at \url{https://www.census.gov/programs-surveys/saipe.html}. In its current form the SAIPE program uses model estimates from the American Community Survey (ACS) 
as the primary data source. Other data sources include Internal Revenue Service (IRS), 
Current Population Survey (CPS), 
and Supplemental Nutrition Assistance Program (SNAP). 

SAIPE production uses a popular small area estimation model to borrow strength from covariates, specifically a univariate FH model on the log-transformed direct ACS 1-year estimates of poverty counts or rates. Covariates at the county level are available to the SAIPE program from ACS, as well as administrative records including tabulations of income tax data obtained from the IRS, and tabulations of participant counts from the SNAP 
\citep[see][for further details]{Bell2007}.  Since SAIPE data are not direct estimates but are actually model estimates based on ACS data, we use ACS 2019 5-year estimates at the county level for the purpose of this application.     

Using our proposed HGT-SME framework, 
we model the ACS 2019 5-year period estimates of county poverty counts of school-aged children in the states of Idaho, Montana, Oregon, and Washington. There are a total of $175$ counties in these states. Our count valued response is the ACS 2019 5-year period estimates of counts of the population under 18 years of age that are below the poverty threshold. We use two error-prone covariates, the log of ACS 2019 5-year period estimates of median income per household with one or more children under 18 years of age and the logit transformed ACS 2019 5-year period estimates of the proportion of the county population that receives SNAP. The log and logit transformations are used to satisfy model assumptions like Gaussianity. We assume these covariates to be independent of each other and hence the priors on them would represent two independent univariate CAR priors. We use the ACS direct estimates for the sampling error as the measurement error variance and use appropriate delta method transformations to adjust for the transformations of the covariates in our model specification. 

In this application, the matrix $\mathbf{S}$ includes an intercept, and a fixed covariate, that we assume have not been measured with error. As this fixed covariate we use the county ‘child tax filer rate’, which is defined as the number of child exemptions in the county claimed on tax returns divided by the county population age $0-17$ years. We adopt this from \cite{Arima2017} and is derived from public use data released by the IRS under its Statement of Income (SOI) program. We set $r=67$ which corresponds to $95\%$ of the positive eigenvalues of the MI operator.  

\cite{Arima2017} in their multi-year analysis, found that the fixed covariate `child tax filer rate' may not be a significant predictor for some years. We thus considered two versions of the HGT-SME model for this application, Model 1 where this fixed covariate is included in the $\mathbf{S}$ matrix and Model 2 where this covariate is excluded. In both versions, a fixed intercept is included in $\mathbf{S}$. We used deviance information criteria \citep[DIC, ][]{Spiegelhalter2002,Celeux2006} and Watanabe Akaike information criteria \citep[WAIC, ][]{Watanabe2010, Gelman2014} to compare the two versions. We also compare the point estimates of the coefficient parameters for the two versions in Table~\ref{table: RD1}. We see that, Model 2 that excludes the fixed covariate has a lower DIC and WAIC compared to Model 1. We also note that the point estimates of the coefficient parameters are not substantially different between the two models, and in fact the $95\%$ credible interval of the coefficient for the fixed covariate, `child tax filer rate' in Model 1 suggests that this covariate might not be a significant predictor for this application. Thus, while similar SAIPE motivated applications uses this variable as an error free covariate, this comparative study suggests Model 2 is the preferred model for this SAIPE motivated illustration and hence for the rest of this section our discussion is based on Model 2.

\begin{table}[t]
\centering 
\caption{Point estimates of the coefficient parameters along with the $95\%$ CI, and the DIC and WAIC values for the two versions of the HGT-SME model considered in the example. $\textcolor{red}{*}$ refers to the case where the $95\%$ CI include zero. \label{table: RD1}}
\resizebox{\columnwidth}{!}{%
\renewcommand{\arraystretch}{1.2}
\setlength{\tabcolsep}{10pt}
\begin{tabular}{ccc|c|c|c|c|c}
& & \multicolumn{1}{ c }{Intercept} & \multicolumn{1}{ c }{$\widehat{\beta}_1$} & \multicolumn{1}{ c }{$\widehat{\beta}_2$}  & \multicolumn{1}{ c }{$\widehat{\delta}_1$} & \multicolumn{1}{ c }{DIC} & \multicolumn{1}{ c }{WAIC} \\ \toprule
\multicolumn{1}{ l | }{\multirow{2}{*}{Model 1 }} & \multicolumn{1}{l|}{ Naive } & -17.71 (-27.23, -7.91) & 2.43 (1.48, 3.37) & 1.51 (1.21, 1.79) & 0.011 (-0.015, 0.036)$^{\textcolor{red}{*}}$ & \\ 
\multicolumn{1}{ l | }{} & \multicolumn{1}{l|}{HGT-SME} & -19.25 (-27.23, -9.81) & 2.58 (1.67, 3.38) & 1.58 (1.26, 1.95) & 0.019 (-0.012, 0.044)$^{\textcolor{red}{*}}$ & 6738.897 & 1604.538 \\
\midrule
\multicolumn{1}{ l | }{\multirow{2}{*}{Model 2 }} & \multicolumn{1}{l|}{ Naive } & ~-18.68 (-27.98,  -9.64) & 2.58 (1.75, 3.44) & 1.54 (1.26, 1.82) & & &  \\ 
\multicolumn{1}{ l | }{} & \multicolumn{1}{l|}{HGT-SME} & ~-20.47 (-28.62,  -11.61) & 2.78 (1.95, 3.55) & 1.63 (1.34, 1.95) & & 4051.873  & 254.8941  \\
\bottomrule
\end{tabular}%
}

\end{table}

\begin{figure}[!htb]
\begin{subfigure}{\textwidth}
\begin{adjustwidth}{-0.5cm}{}
 \centering
    \includegraphics[scale=0.5]{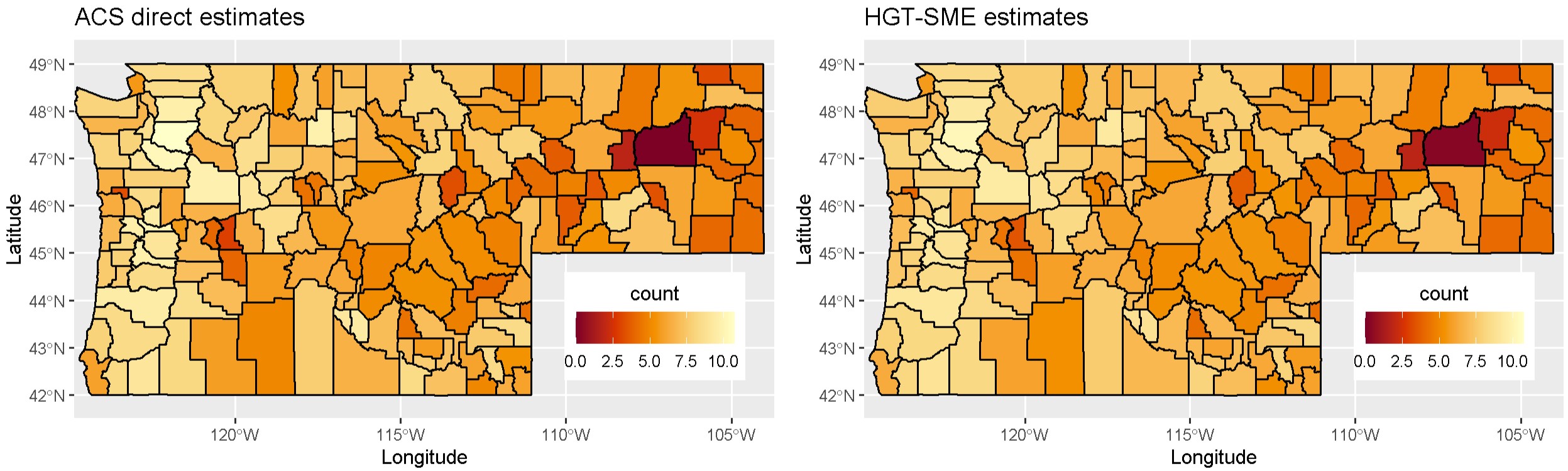}
 \end{adjustwidth} 
\caption{\label{fig:RD1a}}
\end{subfigure}

    \hspace*{\fill}
\begin{subfigure}{\textwidth}
\begin{adjustwidth}{-0.5cm}{}
 \centering
    \includegraphics[scale=0.6]{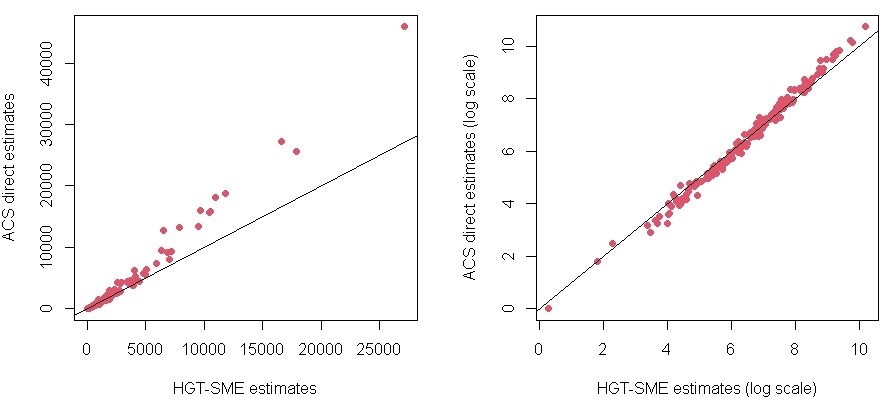}
\end{adjustwidth} 
 \caption{\label{fig:RD1b}}
 \end{subfigure}
\caption{(a) Maps showing the ACS 2019 5-year period estimates of count of county population below poverty in the 4 states and the HGT-SME model estimates of the counts on the log scale. (b) Scatter plot comparing the HGT-SME model estimates against the ACS 2019 5-year period direct estimates of county population under $18$ years in poverty on the original scale and the log scale.  
\label{fig: RD1}}
\end{figure}

We now compare the HGT-SME model estimates with the ACS direct estimates. Figure~\ref{fig:RD1a} maps the ACS 2019 5-year estimates of the count of county population under 18 years of age that are below the poverty threshold and the HGT-SME model estimates of the same on the log scale. From the plots, we see that our model estimates are able to capture the spatial trend in the ACS direct estimates. For the lower count counties, the model estimates trace the direct estimates closely, and for the higher count counties, the model estimates exhibit shrinkage from the extreme value while still following the spatial trend of the ACS direct estimates. Figure~\ref{fig:RD1b} compares the HGT-SME model estimates against the ACS direct estimates and supports the shrinkage of the extremely high counts that we noted previously. Note the difference in ranges of the $x-$ and $y-$axis. This shrinkage is expected as the higher count counties report higher measurement error and, hence, the direct estimates are noisier.

Figure~\ref{fig: RD1.2} compares the HGT-SME model estimates with estimates from the naive model. Figure~\ref{fig:RD1.2a} shows that the HGT-SME model estimates deviate from the naive model estimates which suggest that accounting for the measurement error does have an affect in the small area estimations. Figure~\ref{fig:RD1.2a} also confirms that while the HGT-SME estimates and the naive model estimates are close, for higher count counties, i.e. counties that report extreme values, the HGT-SME model produces closer estimates to the ACS direct estimates compared to the naive model. Figure~\ref{fig:RD1.2b} shows that the naive model does report higher posterior standard errors. The effect of the measurement error may not be too pronounced, but for a different choice of covariates and response, perhaps one not requiring a log or logit transformation, the effects may be  more pronounced. We also note here that the effect of measurement error could also be obscured by the other covariates considered here. 

\afterpage{%
\begin{figure}[t]
\begin{subfigure}{0.5\columnwidth}
    \includegraphics[scale=0.5]{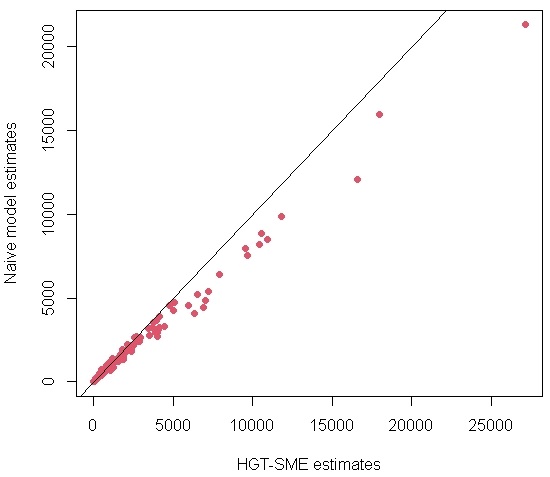}
    \caption{\label{fig:RD1.2a}}
\end{subfigure}%
\hspace*{\fill}
\begin{subfigure}{0.5\columnwidth}
    \includegraphics[scale=0.5]{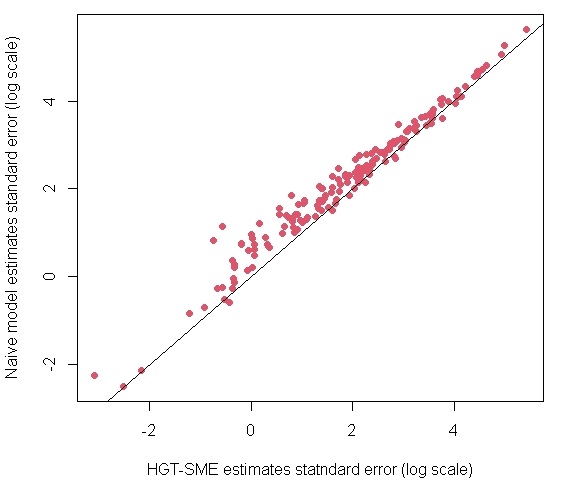}
    \caption{\label{fig:RD1.2b}}
\end{subfigure}
\caption{(a) Plot comparing the HGT-SME model estimates against the naive model. (b) Plot comparing the posterior standard error in the HGT-SME model estimates against the standard error of the naive model.\label{fig: RD1.2}}
\end{figure}
}
Figure~\ref{fig: RD2} shows the ACS standard errors per county in the four states. We also plot the standard errors of the HGT-SME model estimates as well as the naive model. We see that the model estimates always have lower standard errors than the direct estimates and for some of the counties, this difference can be quite large as seen in the plots showing the percentage reduction in standard error.  We also plot the relative percentage reduction in standard error which tells us how much  the HGT-SME model reduced the standard error in the estimates relative to the naive model. We see that for some counties, the relative percent reduction is negative suggesting the naive model performs better than the HGT-SME model in these counties, but for a majority of the counties, the HGT-SME model shows an approximate $10-25\%$ relative reduction in standard error over the naive model, suggesting that the measurement error assumption in the model specification, in fact, is beneficial to the estimation process. This observation also aligns with what we saw in the empirical simulation studies where the pseudo data generated very closely resembled the real data used here. 

\afterpage{%
\begin{landscape}
\begin{figure}[!htb]
    \centering
    \includegraphics[scale=0.425]{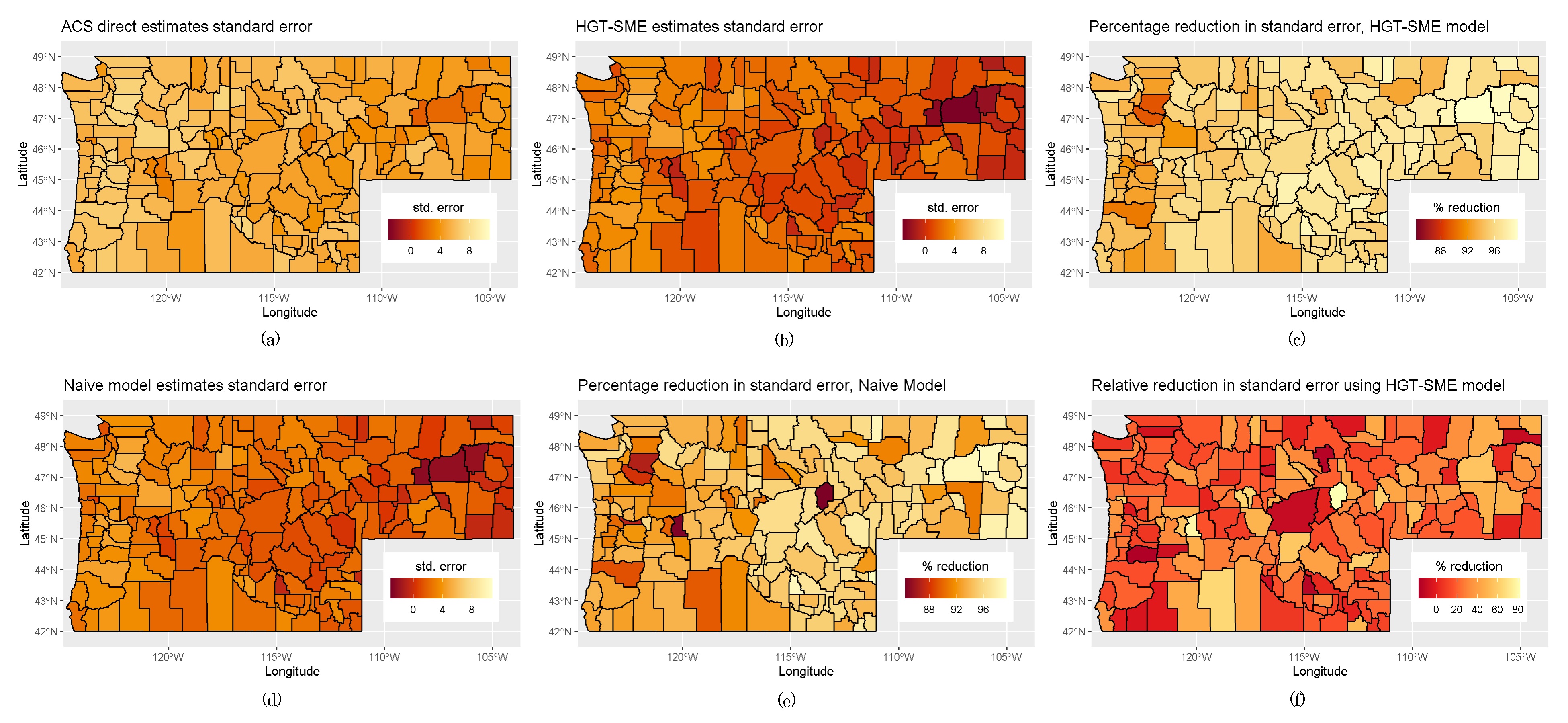}
\caption{(a) Map of the study area showing the standard error in the ACS 2019 direct estimates on the log scale. (b-c) Map of the study area showing the standard error in the HGT-SME model estimates and the percentage reduction in standard error achieved by the HGT-SME model over the ACS direct estimates. (d-e) Map of the study area showing the standard error in the naive model estimates and the percentage reduction in standard error achieved by the naive model over the ACS direct estimates. (f) Map of the study area showing the relative percentage reduction  in standard error achieved by the HGT-SME model estimates over the naive model. \label{fig: RD2}}
\end{figure}
\end{landscape}
}
\subsection{Joint Estimation of Housing Cost and Count of Population Below Poverty in Pacific NW}

The U.S. Census Bureau and the U.S. Department of Housing and Urban Development (HUD) prepare the Comprehensive Housing Affordability Strategy (CHAS) database to demonstrate the number of households in need of housing assistance and to serve as the strategic guide for housing and community development activities for low and moderate-income households \citep{Hoben1992}. The CHAS data combine ACS microdata with HUD-adjusted Median Family Incomes (HAMFIs) to create estimates of the number of households that qualify for HUD assistance. 
HUD uses ACS 5-year data to ascertain the Difficult Development Areas (DDAs) and the Qualified Census Tracts (QCTs) which are used in the Low-Income Housing Tax Credit (LIHTC) program to allocate federal funds, tax credits, and investment opportunities in low income areas. 

Motivated by these uses of ACS data, we use our proposed HGT-SME framework to jointly model ACS 2019 5-year estimates of median housing costs and county of population below poverty threshold. This enables us to illustrate the use of our proposed framework by jointly modeling both a Gaussian and a non-Gaussian response type dataset. We treat the log of ACS 2019 5-year period estimates of median housing cost per county in the states of Idaho, Montana, Oregon, and Washington as the continuous Gaussian response and the ACS 2019 5-year period estimates of counts of county population below the poverty threshold in these states as the count-valued Poisson response. Similar to the single response example, we use two error prone covariates - the log of ACS 2019 5-year period estimates of median household income per county in these states and the logit transformed ACS 2019 5-year period estimates of the proportion of county population that receives SNAP.  
We use the ACS direct estimates for the sampling error as the measurement error variance for the covariates and use appropriate delta method transformations to adjust for the transformations of the covariates in our model specification.


Based on our observations in the single type response example in Section~\ref{Section: RealEg1}, we decided to include only a fixed intercept in the $\mathbf{S}$ matrix. However, as we have exhibited in Section~\ref{Section: RealEg1} the model specification can handle multiple fixed covariates and can be included if it is found significant for a given data application.

\begin{figure}[!htb]
\begin{subfigure}{\textwidth}
    \centering
    \includegraphics[scale=0.55]{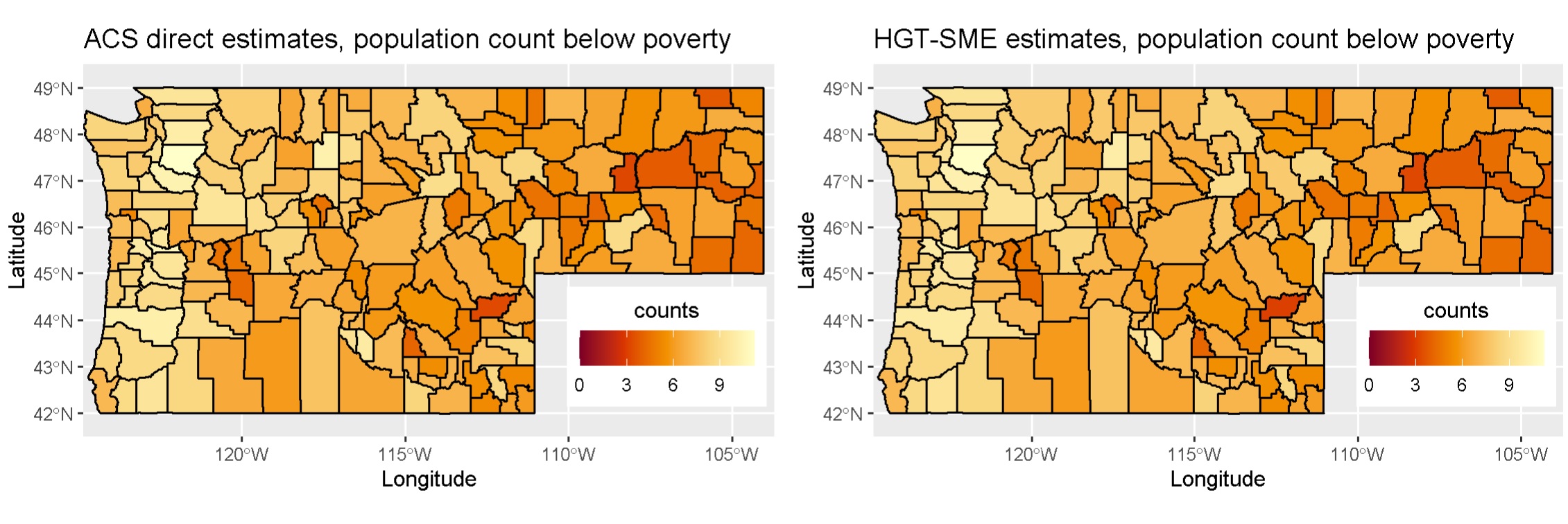}
\caption{\label{fig:RD_mult_1a}}
\end{subfigure}
    \hspace*{\fill}
\begin{subfigure}{\textwidth}
\centering
\includegraphics[scale=0.55]{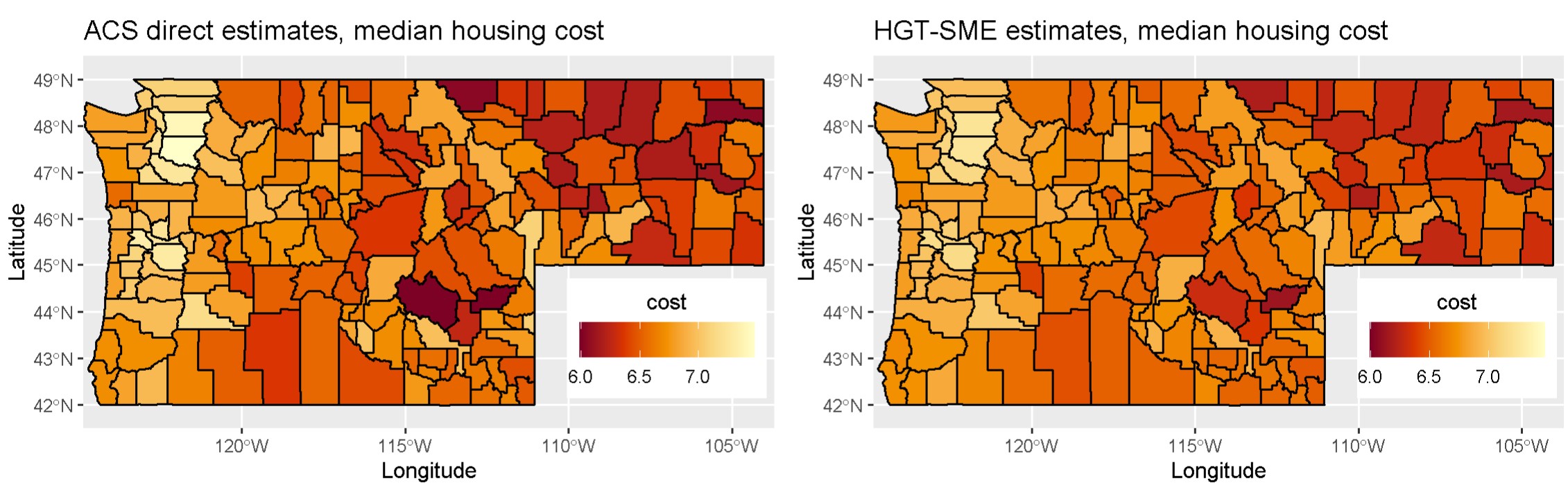}
 \caption{\label{fig:RD_mult_1b}}
 \end{subfigure}
\caption{(a) Maps showing the ACS 2019 5-year period estimates of count of county population below the poverty threshold in the 4 states and the HGT-SME model estimates of the counts, both on the log scale. (b) Maps showing the log of ACS 2019 5-year period estimates of median housing cost in the 4 states and the HGT-SME model estimates of the log median housing costs.
\label{fig: RD3_est}}
\end{figure}

\begin{figure}[!htb]
    \centering
    \includegraphics[scale=0.3]{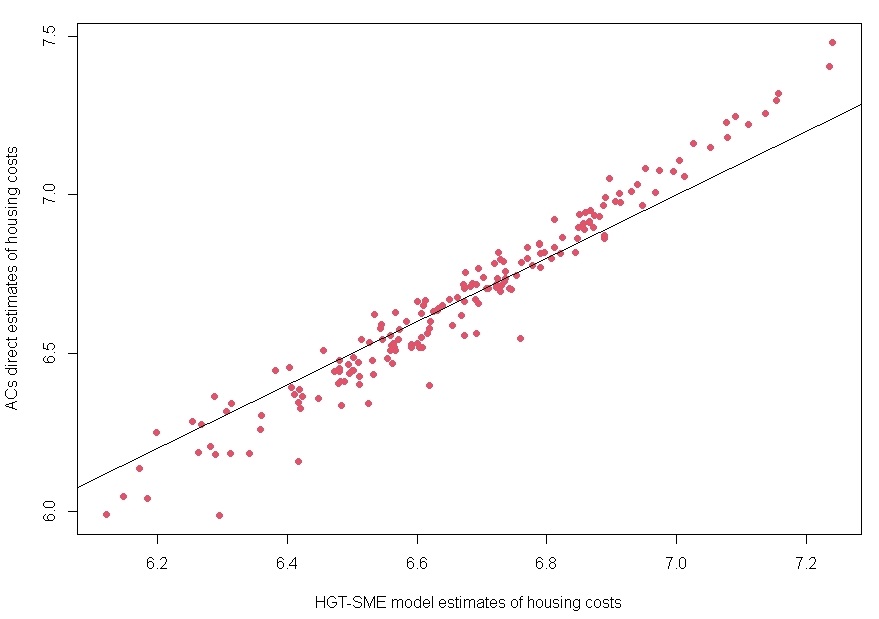}
    \includegraphics[scale=0.3]{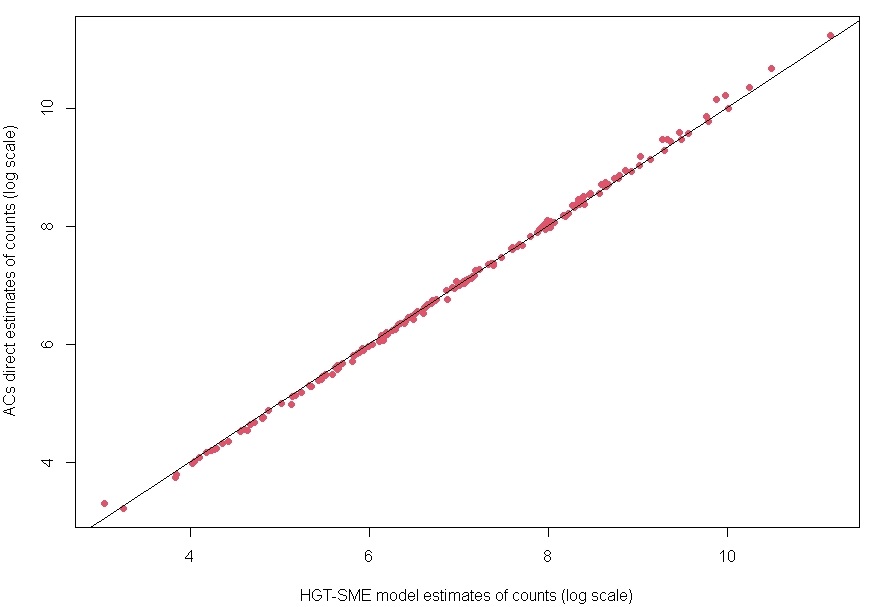}
    \caption{Scatter Plots comparing the ACS 2019 direct estimates of the Gaussian (left) and Poisson (right) Responses against the HGT-SME model estimates. \label{fig: RD_mult_1c}}
\end{figure}

We compare the ACS 2019 direct estimates with the HGT-SME model estimates in Figure~\ref{fig: RD3_est}. Figure~\ref{fig:RD_mult_1a} compares the ACS 2019 direct estimates of the count of county population below the poverty threshold against the HGT-SME model estimates of the count, both on the log scale. We see that the model estimates are able to fully capture the spatial distribution and resemble the ACS direct estimates to a great extent. When we compare these estimates with our example in Section~\ref{Section: RealEg1}, we see that the model estimates for the Poisson response actually benefited from the joint modelling as suggested by the scatter plot in Figure~\ref{fig: RD_mult_1c}, where we observe that even though the HGT-SME estimates still suffer shrinkage at the extreme values, the shrinkage is much less compared to what we observed in the Section~\ref{Section: RealEg1} (Figure~\ref{fig:RD1b}). Although a direct comparison between the two examples is not appropriate, the observation about the shrinkage stands nevertheless. Figure~\ref{fig:RD_mult_1b} compares the ACS 2019 direct estimates of log median housing cost per county against the HGT-SME estimates of the cost. Here, too, we observe that the model estimates are successful in capturing the spatial distribution, and the model estimates closely trace the ACS estimates as suggested by the scatter plot in Figure~\ref{fig: RD_mult_1c}. 

\begin{figure}[t]
\begin{subfigure}{\textwidth}
\centering
\includegraphics[scale=0.3]{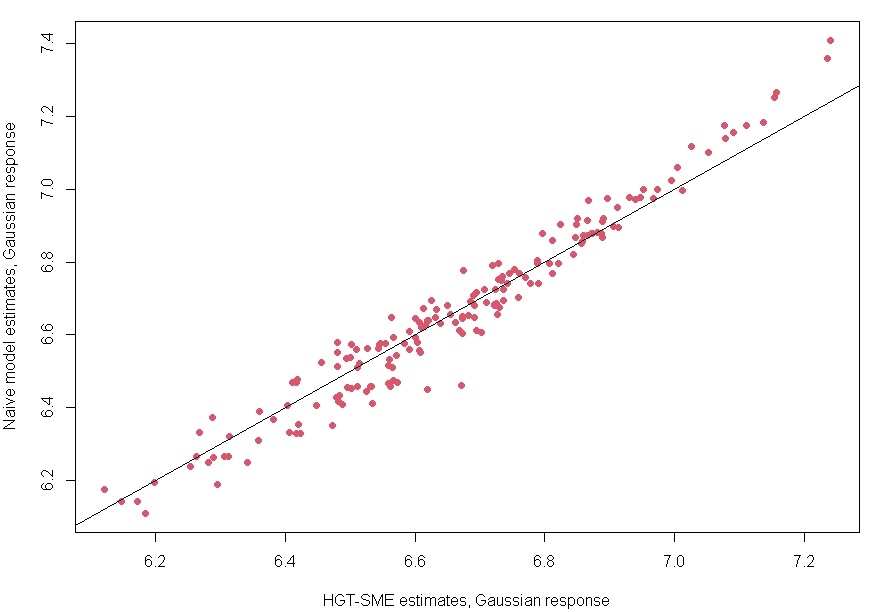}
\includegraphics[scale=0.3]{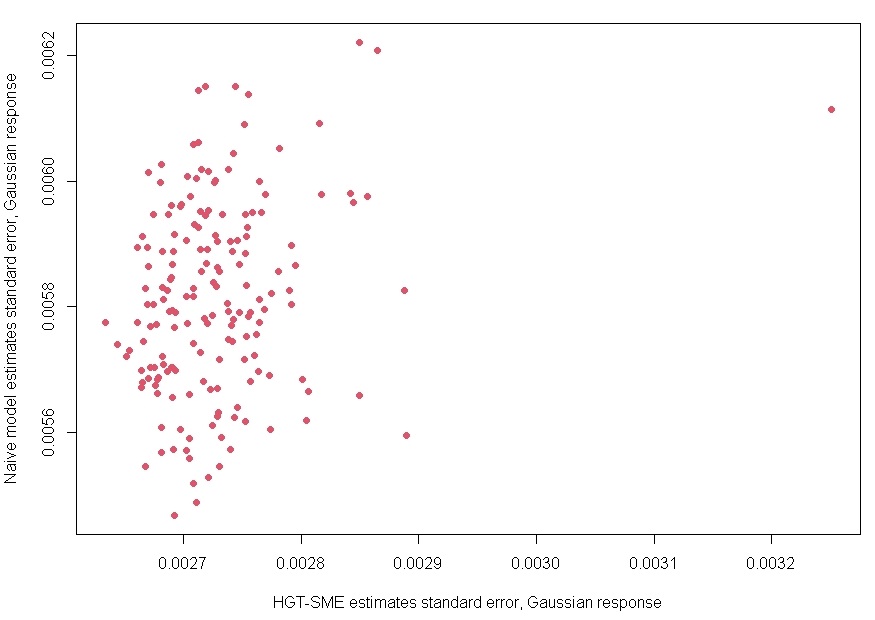}
\caption{\label{fig:RD_mult_2a}}

\end{subfigure}

\begin{subfigure}{\textwidth}
\centering
\includegraphics[scale=0.3]{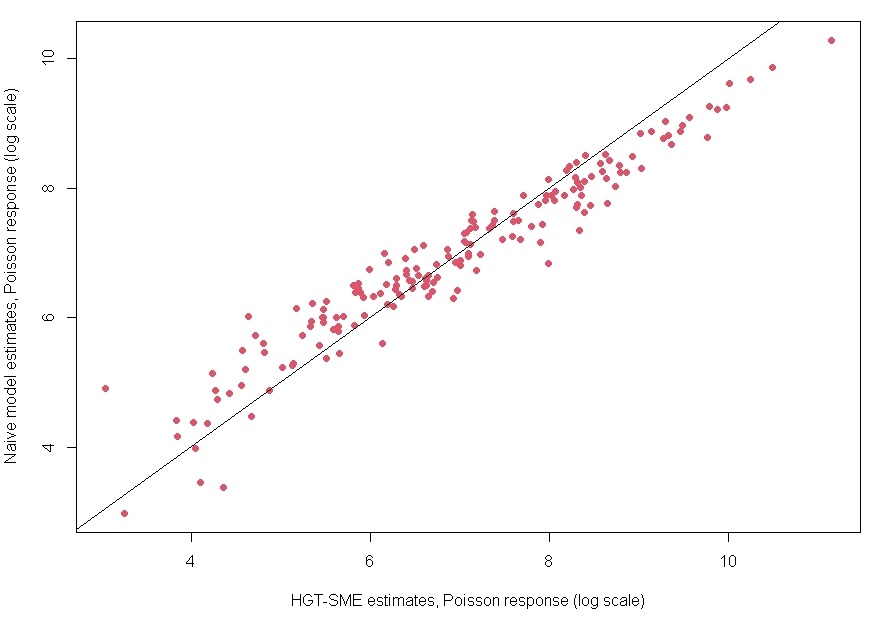}
\includegraphics[scale=0.3]{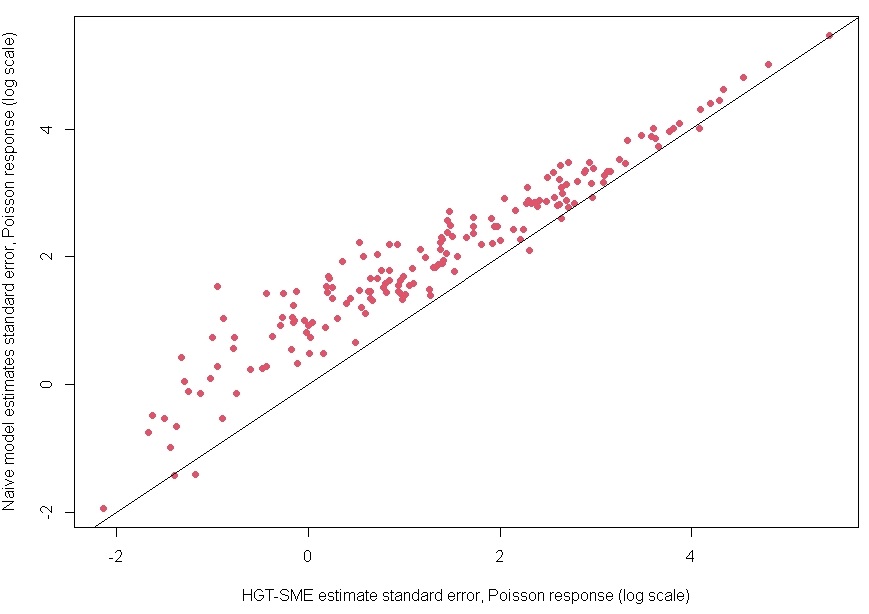}
\caption{\label{fig:RD_mult_2b}}
\end{subfigure}
\caption{Scatter Plots comparing the HGT-SME model estimates against the naive model estimates, and the corresponding standard error in these estimates of: (a) the log median housing cost per county, (b) the county population count below poverty threshold.}
\end{figure}

Scatter Plots in Figure~\ref{fig:RD_mult_2a} and \ref{fig:RD_mult_2b} compare the HGT-SME model estimates against the naive model estimates as well as the corresponding standard errors in the model estimates on the log scale. Both the plots show that assuming the measurement error in the model specification does impact the model estimates. While the impact is slightly nuanced in the estimates for Gaussian response case which can be attributed in part to the log transformation that was required to satisfy the Gaussianity assumption, but the effect is quite pronounced for the Poisson response as can be seen from Figure~\ref{fig:RD_mult_2b}. The posterior standard errors too are impacted by the measurement error assumption as we see for both cases, the HGT-SME model estimates have much lower uncertainty as compared to the naive model estimates. 

Figures~\ref{fig: RD_mult_3a} and \ref{fig: RD_mult_3b} map the reported ACS 2019 direct estimates standard errors against the standard errors in HGT-SME and the naive model estimates for both the Gaussian and Poisson response cases. First, we observe that, for either type of response, the model estimates exhibit a significant reduction in standard errors from the ACS estimates. We also map the percentage reduction in standard error that the HGT-SME model achieves over the ACS estimates. We see that for both response types, the percent reduction in standard error using the HGT-SME model is substantial. 
This observation also aligns with the analysis results in \cite{Arima2017}, where the authors note that the direct survey estimates can `have standard errors up to about seven times larger than those of the model estimates.' They also note that as the sample size increases, the difference between the direct estimates and the model's standard error decreases. We observe something similar, where in Figures~\ref{fig: RD_mult_3a} and \ref{fig: RD_mult_3b}, the counties with the lowest reduction in standard error for both the HGT-SME and naive models (marked in darker shades) represent some of the most populous counties in the study area, around cities like Seattle, WA, and Portland, OR. If we compare the percentage reduction in standard error,  we observe that our proposed HGT-SME model outperforms the naive model suggesting that the measurement error assumption helps in the estimation process. We see from Figures~\ref{fig: RD_mult_3a} and \ref{fig: RD_mult_3b}, that using our proposed model, we obtain a significantly high relative reduction in standard errors over the naive model, suggesting that the measurement error assumption also reduces the model uncertainty.

\afterpage{%
\begin{landscape}
\begin{figure}[!htb]
    \centering
    \includegraphics[scale=0.425]{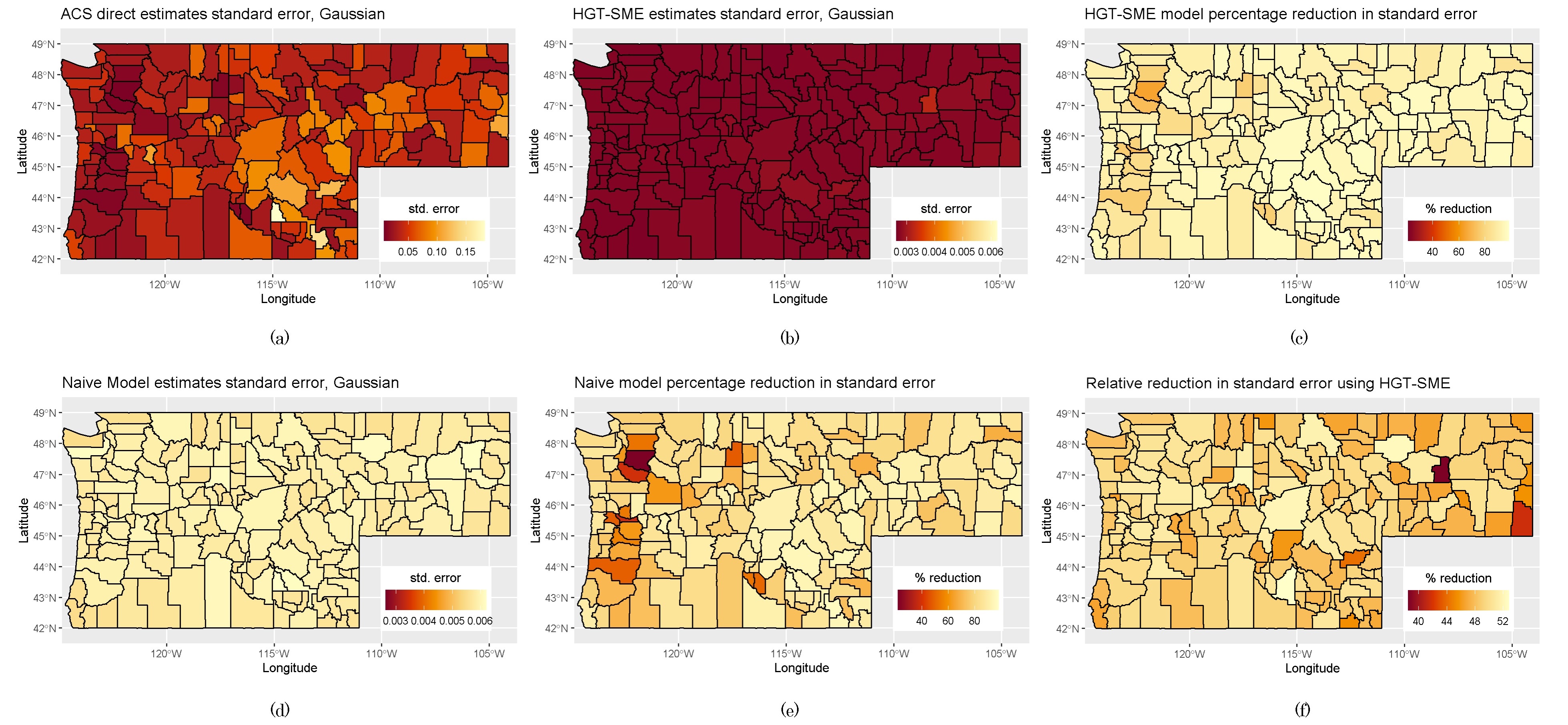}
\caption{(a) Map of the study area showing the standard error in the ACS 2019 direct estimates of log median housing cost. (b-c) Map of the study area showing the standard error in the HGT-SME model estimates and the percentage reduction in standard error achieved by the HGT-SME model over the ACS direct estimates. (d-e) Map of the study area showing the standard error in the naive model estimates and the percentage reduction in standard error achieved by the naive model over the ACS direct estimates. (f) Map of the study area showing the relative percentage reduction in standard error achieved by the HGT-SME model estimates over the naive model. Maps (a), (b), and (d) do not share the same scale in legend due to the significant difference in the range of the corresponding values. \label{fig: RD_mult_3a}}
\end{figure}

\begin{figure}[!htb]
    \centering
    \includegraphics[scale=0.425]{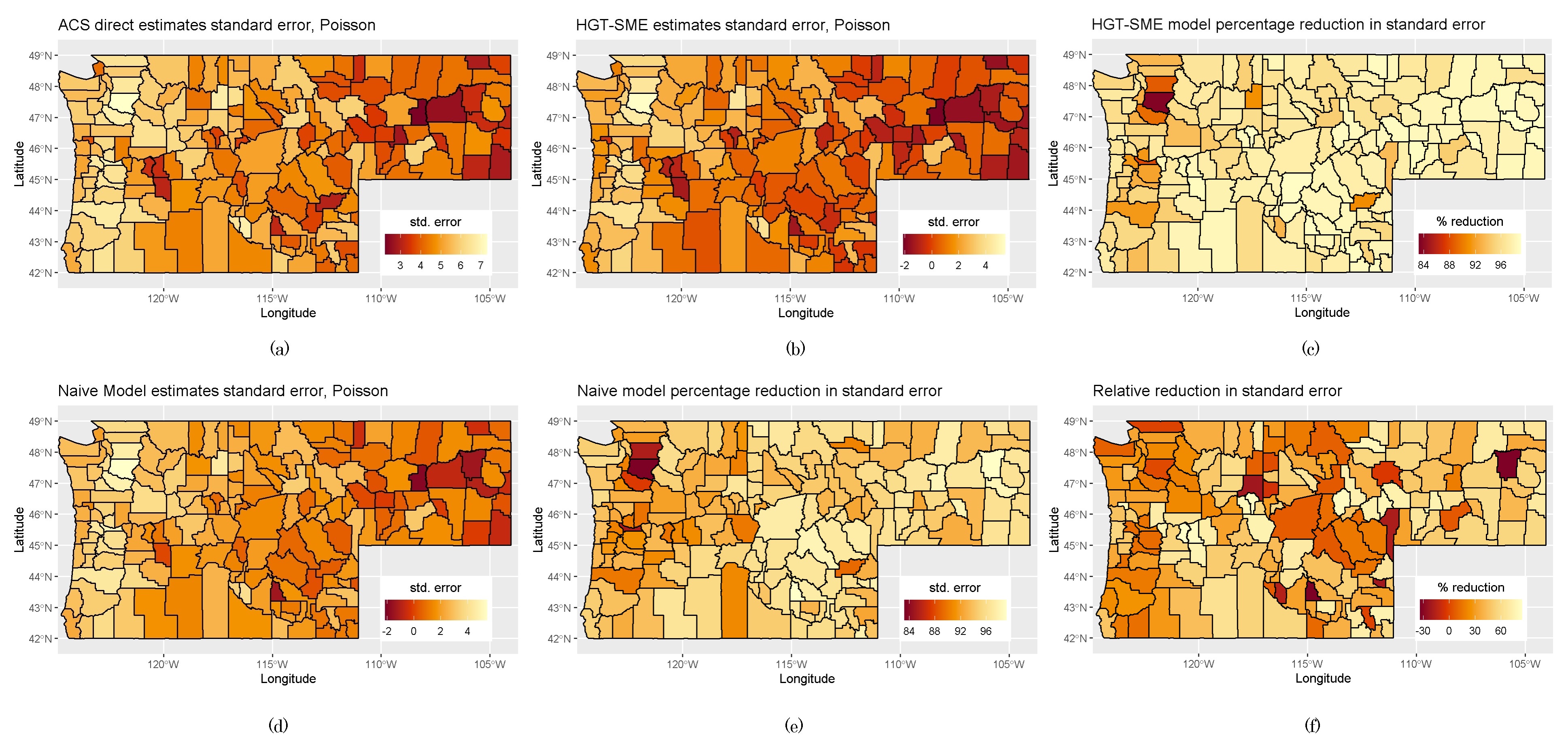}
\caption{(a) Map of the study area showing the standard error in the ACS 2019 direct estimates of log median housing cost. (b-c) Map of the study area showing the standard error in the HGT-SME model estimates and the percentage reduction in standard error achieved by the HGT-SME model over the ACS direct estimates. (d-e) Map of the study area showing the standard error in the naive model estimates and the percentage reduction in standard error achieved by the naive model over the ACS direct estimates. (f) Map of the study area showing the relative percentage reduction in standard error achieved by the HGT-SME model estimates over the naive model.\label{fig: RD_mult_3b}}
\end{figure}
\end{landscape}
}
\section{Discussion} \label{Section: Discussion}
In this paper, we have proposed a fully Bayesian approach to model spatially distributed survey data which can be Gausian or non-Gaussian. Our proposed framework is general enough to be extended to the spatio-temporal paradigm as well and is the subject of future research. We have used the HGT component to deal with the computational complexity associated with modeling non-Gaussian data. Following \cite{Bradley2021} the HGT component helps provide a suitable transformation for non-Gaussian data like count-valued Poisson data or integer-valued Binomial data. The HGT component of our model also enables us to jointly model multi-type response datasets under a unified framework without any significant changes to the model hierarchy.

Our model specification allows one to incorporate prior knowledge of the spatial process into the HGT-SME model hierarchy and is flexible enough to allow different spatial structures on the covariates depending upon the data examples and model applications. To model the spatial dependency in the survey data, we assumed a CAR model on the covariates. We expect a seamless adaptation of other spatial structures in the model hierarchy. The strength of the spatial association can be controlled by a suitable prior on the spatial correlation factor $\rho$. 
Another aspect of our proposed model is that it deals with the measurement error that is often associated with survey data. The proposed model assumes the presence of measurement error in the covariates and corrects for the same by considering a classical measurement error model. The framework also includes a fixed effect component which comprises covariates that we assume have no measurement error. The assumption of independence between the fixed covariates and the error-prone covariates is a potential limitation and may motivate future work. 

The HGT-SME model is flexible enough for several different specifications, including different classes of basis functions. We use MI basis function expansion primarily for illustrations, in part motivated by its dimension reduction properties. However, the choice can be viewed as problem specific. Determining the dimension of the basis functions matrix is an open problem and often application specific \citep[see][]{Bradley2011}. For our data applications and simulation studies, we have relied on sensitivity analysis for this choice. 

We have illustrated our model performance by designing empirical simulation studies based on data calibrated towards survey data produced by the ACS. We have presented simulation studies for both single-type responses as well as multi-type responses. From these studies, we have observed that by accounting for the measurement error in our model specification, we can reduce the mean squared prediction error and the absolute bias in the small area estimates over the naive model which is the case where we ignore the measurement error in survey estimates. 

We have motivated our model applications using two application from the ACS. In the first example, we calibrated our model using a SAIPE data example. In this example, we observed our proposed model produced significantly lower standard errors associated with the small area estimates. In the second example, we have exhibited a multi-type response application, where we jointly estimated median housing costs and count of population below poverty, thus a joint modeling of a Gaussian and Poisson response under a unified latent process framework. 
There too we have shown that our proposed model produces significantly lower standard error with the area estimates. The joint estimation also showed higher relative reduction in posterior standard error in the small area estimates using the proposed HGT-SME framework over the naive model of ignoring measurement error in covaiates. 

Future work can be motivated by exploring different measurement error model formulations (e.g., the Berkson error model) as well as different spatial structure on the error prone covariates. We have assumed independent univariate CAR priors on error prone covariates, which might not be realistic in case of survey data like ACS and hence extension to MCAR specification could constitute future research. A potential extension to spatio-temporal data could also be explored.

\section*{Acknowledgements}

This research was partially supported by the U.S.~National Science Foundation (NSF) under NSF grants SES-1853096 and NCSE-2215168. This article is released to inform interested parties of ongoing research and to encourage discussion. The views expressed on statistical issues are those of the authors and not those of the NSF or U.S. Census Bureau.\vspace*{-8pt}
  

\section*{Appendix}

\subsection*{Full Conditional Distributions}
\label{FullConditionals}
The full conditional distributions for the various components in the proposed HGT-SME model are discussed here. 
\subsubsection*{Random Effect Parameters}
For the error prone covariate, $\mathbf{W}$, we only present the full conditional distribution for the univariate case, i.e. $p=1$. Depending upon which specification for an MCAR model is used in the model, the full conditional will change. 
\begin{align*}
    \mathbf{W}|\cdot &\propto \exp{\left \{ -\frac{1}{2\tau^2} \left ( \mathbf{h}- \mathbf{W}\boldsymbol{\beta } -\mathbf{S}\boldsymbol{\delta}-\mathbf{M}\boldsymbol{\eta}-\boldsymbol{\xi}\right )^2 \right \}}\times \exp{ \left \{ -\frac{1}{2} (\mathbf{X}-\mathbf{W})^{'}\boldsymbol{\Sigma}_{\mathbf{U}}^{-1}(\mathbf{X}-\mathbf{W})\right \} }\\
    &\times \exp{ \left \{ -\frac{1}{2} (\mathbf{W}-\mathbf{0})^{'}(\sigma^2_{\boldsymbol{W}}\boldsymbol{\Sigma}_{\mathbf{W}})^{-1}(\mathbf{W}-\mathbf{0})\right \} }\\
    &= \exp{\left \{ - (\mathbf{W}-\boldsymbol{\mu}_{\mathbf{W}}^{*})^{'} (\boldsymbol{\Sigma}_{\mathbf{W}}^{*})^{-1} (\mathbf{W}-\boldsymbol{\mu}_{\mathbf{W}}^{*}) \right \} }.
\end{align*}
Hence,
$\mathbf{W}|\cdot \sim \hbox{Gaussian}(\boldsymbol{\mu}_{\mathbf{W}}^{*},\boldsymbol{\Sigma}_{\mathbf{W}}^{*}), $\\
where $\boldsymbol{\Sigma}_{\mathbf{W}}^{*}=\left ( \frac{\boldsymbol{\beta}^{'}\boldsymbol{\beta}}{\tau^2}\mathbf{I}_N +\frac{1}{\sigma^2_{\mathbf{W}}}\boldsymbol{\Sigma}_{\mathbf{W}}^{-1}+\boldsymbol{\Sigma}_{\mathbf{U}}^{-1} \right )^{-1}$, and  $\boldsymbol{\mu}_{\mathbf{W}}^{*}=\boldsymbol{\Sigma}_{\mathbf{W}}^{*} \left ( \boldsymbol{\Sigma}_{\mathbf{U}}^{-1} \mathbf{X}  + \frac{\boldsymbol{\beta}}{\tau^2}\left ( \mathbf{h}-\mathbf{S}\boldsymbol{\delta} - \mathbf{M}\boldsymbol{\eta}-\boldsymbol{\xi}\right ) \right )$.

\begin{align*}
    \boldsymbol{\eta}|\cdot &\propto \exp{\left \{ -\frac{1}{2\tau^2} \left ( \mathbf{h}- \mathbf{W}\boldsymbol{\beta} -\mathbf{S}\boldsymbol{\delta}-\mathbf{M}\boldsymbol{\eta}-\boldsymbol{\xi}\right )^2 \right \}}\times \exp{\left \{ -\frac{1}{2} (\boldsymbol{\eta}-\mathbf{0})^{'} (\sigma^2_{\boldsymbol{\eta}}\mathbf{I}_r)^{-1}(\boldsymbol{\eta}-\mathbf{0}) \right \} }\\
    &= \exp{ \left \{ -\frac{1}{2}(\boldsymbol{\eta}-\boldsymbol{\mu}_{\boldsymbol{\eta}}^{*})^{'}\Sigma_{\boldsymbol{\eta}}^{-1}(\boldsymbol{\eta}-\boldsymbol{\mu}_{\boldsymbol{\eta}}^{*})\right \}},
\end{align*}
Hence, $\boldsymbol{\eta}|\cdot \sim \hbox{Gaussian}\left ( \boldsymbol{\mu_{\boldsymbol{\eta}}}^{*},\Sigma_{\boldsymbol{\eta}}  \right ),$\\ where  $\Sigma_{\boldsymbol{\eta}}=\left ( \frac{1}{\sigma^2_{\boldsymbol{\eta}}}\mathbf{I}_r+\frac{1}{\tau^2} \mathbf{M}^{'}\mathbf{M}\right )^{-1}$ and $\boldsymbol{\mu_{\boldsymbol{\eta}}}^{*}=\frac{1}{\tau^2}\Sigma_{\boldsymbol{\eta}}\mathbf{M}^{'}\left ( \mathbf{h}- \mathbf{W}\boldsymbol{\beta} -\mathbf{S}\boldsymbol{\delta}-\boldsymbol{\xi}\right )$.

\begin{align*}
    \boldsymbol{\xi}|\cdot &\propto  \exp{\left \{ -\frac{1}{2\tau^2} \left ( \mathbf{h}- \mathbf{W}\boldsymbol{\beta} -\mathbf{S}\boldsymbol{\delta}-\mathbf{M}\boldsymbol{\eta}-\boldsymbol{\xi}\right )^2 \right \}}\times \exp{\left \{ -\frac{1}{2}(\boldsymbol{\xi}-\mathbf{0})^{'} (\sigma^2_{\boldsymbol{\xi}}\mathbf{I}_N)^{-1}(\boldsymbol{\xi}-\mathbf{0}) \right \} }\\
    &= \exp{ \left \{ -\frac{1}{2}(\boldsymbol{\eta}-\boldsymbol{\xi}_{\boldsymbol{\xi}}^{*})^{'}\Sigma_{\boldsymbol{\xi}}^{-1}(\boldsymbol{\xi}-\boldsymbol{\mu}_{\boldsymbol{\xi}}^{*})\right \}},
\end{align*}
Hence, $\boldsymbol{\xi}|\cdot \sim \hbox{Gaussian}\left ( \boldsymbol{\mu_{\boldsymbol{\xi}}}^{*},\Sigma_{\boldsymbol{\xi}}  \right ),$\\
where  $\Sigma_{\boldsymbol{\xi}}=\left ( \frac{1}{\sigma^2_{\boldsymbol{\xi}}}\mathbf{I}_N+\frac{1}{\tau^2} \mathbf{I}_N \right )^{-1}$ and   $\boldsymbol{\mu_{\boldsymbol{\xi}}}^{*}=\frac{1}{\tau^2}\Sigma_{\boldsymbol{\xi}}\left ( \mathbf{h}- \mathbf{W}\boldsymbol{\beta}-\mathbf{S}\boldsymbol{\delta} -\mathbf{M}\boldsymbol{\eta}\right )$.

\subsubsection*{Coefficient Parameter(s)}

\begin{align*}
    \boldsymbol{\beta}|\cdot &\propto \exp{\left \{ -\frac{1}{2\tau^2} \left ( \mathbf{h}- \mathbf{W}\boldsymbol{\beta} -\mathbf{S}\boldsymbol{\delta}-\mathbf{M}\boldsymbol{\eta}-\boldsymbol{\xi}\right )^2 \right \}}\times \exp{\left \{-\frac{1}{2} (\boldsymbol{\beta}-0)^{'}(\sigma^2_{\boldsymbol{\beta}}\mathbf{I}_p)^{-1}(\boldsymbol{\beta}-0) \right \}}\\
    &= \exp{ \left \{ -\frac{1}{2}(\boldsymbol{\beta}-\boldsymbol{\mu}_{\boldsymbol{\beta}}^{*})^{'}\Sigma_{\boldsymbol{\beta}}^{-1}(\boldsymbol{\beta}-\boldsymbol{\mu}_{\boldsymbol{\beta}}^{*})\right \}},
\end{align*}
Hence, $\boldsymbol{\beta}|\cdot \sim \hbox{Gaussian} \left ( \boldsymbol{\mu_{\boldsymbol{\beta}}}^{*},\Sigma_{\boldsymbol{\beta}}  \right ),$\\
where  $\Sigma_{\boldsymbol{\beta}}=\left ( \frac{1}{\sigma^2_{\boldsymbol{\beta}}} \mathbf{I}_p +\frac{1}{\tau^2} \mathbf{W}^{'}\mathbf{W}\right )^{-1}$ and $\boldsymbol{\mu_{\boldsymbol{\beta}}}^{*}=\frac{1}{\tau^2}\Sigma_{\boldsymbol{\beta}}\mathbf{W}^{'}\left ( \mathbf{h}- \mathbf{S}\boldsymbol{\delta}- \mathbf{M}\boldsymbol{\eta} -\boldsymbol{\xi}\right )$.

\begin{align*}
    \boldsymbol{\delta}|\cdot &\propto \exp{\left \{ -\frac{1}{2\tau^2} \left ( \mathbf{h}- \mathbf{W}\boldsymbol{\beta} -\mathbf{S}\boldsymbol{\delta}-\mathbf{M}\boldsymbol{\eta}-\boldsymbol{\xi}\right )^2 \right \}}\times \exp{\left \{-\frac{1}{2} (\boldsymbol{\delta}-0)^{'}(\sigma^2_{\boldsymbol{\delta}}\mathbf{I}_q)^{-1}(\boldsymbol{\delta}-0) \right \}}\\
    &= \exp{ \left \{ -\frac{1}{2}(\boldsymbol{\delta}-\boldsymbol{\mu}_{\boldsymbol{\delta}}^{*})^{'}\Sigma_{\boldsymbol{\delta}}^{-1}(\boldsymbol{\delta}-\boldsymbol{\mu}_{\boldsymbol{\delta}}^{*})\right \}},
\end{align*}
Hence, $\boldsymbol{\delta}|\cdot \sim \hbox{Gaussian} \left ( \boldsymbol{\mu_{\boldsymbol{\delta}}}^{*},\Sigma_{\boldsymbol{\delta}}  \right ),$\\
where  $\Sigma_{\boldsymbol{\delta}}=\left ( \frac{1}{\sigma^2_{\boldsymbol{\delta}}} \mathbf{I}_q +\frac{1}{\tau^2} \mathbf{S}^{'}\mathbf{S}\right )^{-1}$ and  $\boldsymbol{\mu_{\boldsymbol{\delta}}}^{*}=\frac{1}{\tau^2}\Sigma_{\boldsymbol{\delta}}\mathbf{S}^{'}\left ( \mathbf{h}- \mathbf{W}\boldsymbol{\beta}- \mathbf{M}\boldsymbol{\eta} -\boldsymbol{\xi}\right )$.

\subsubsection*{Variance Parameters}

\begin{align*}
    \sigma^2_{\mathbf{W}}|\cdot &\propto  (\sigma^2_{\mathbf{W}})^{-\frac{N}{2}}\exp{ \left \{ -\frac{1}{2} \mathbf{W}^{'}(\sigma^2_{\boldsymbol{W}}\boldsymbol{\Sigma}_{\mathbf{W}})^{-1}\mathbf{W}\right \} }\times (\sigma^2_{\mathbf{W}_i})^{-\alpha_{\mathbf{W}_i}-1} \exp{\left \{ -\frac{\beta_{\mathbf{W}_i}}{\sigma^2_{\mathbf{W}_i}} \right \} } I(\sigma^2_{\mathbf{W}}>0)\\
    &= (\sigma^2_{\mathbf{W}})^{-\alpha_{\mathbf{W}}^{*}-1} \exp{\left \{ -\frac{\beta_{\mathbf{W}}^{*}}{\sigma^2_{\mathbf{W}}} \right \} } I(\sigma^2_{\mathbf{W}}>0),
\end{align*}
Hence,
$\sigma^2_{\mathbf{W}}|\cdot \sim \hbox{IG} (\alpha_{\mathbf{W}}^{*},\beta_{\mathbf{W}}^{*}) I(\sigma^2_{\mathbf{W}}>0),$ where $\alpha_{\mathbf{W}}^{*}=\alpha_{\mathbf{W}}+\frac{N}{2}$, and $\beta_{\mathbf{W}}^{*}=\beta_{\mathbf{W}}+0.5(\mathbf{W}^{'}(\boldsymbol{\Sigma}_{\mathbf{W}})^{-1}\mathbf{W})$. 

\begin{align*}
    \sigma^2_{\boldsymbol{\eta}}|\cdot &\propto  (\sigma^2_{\boldsymbol{\eta}})^{-\frac{r}{2}}\exp{ \left \{ -\frac{1}{2} \boldsymbol{\eta}^{'}(\sigma^2_{\boldsymbol{\eta}}\mathbf{I}_r)^{-1}\boldsymbol{\eta}\right \} }\times (\sigma^2_{\boldsymbol{\eta}})^{-\alpha_{\boldsymbol{\eta}}-1} \exp{\left \{ -\frac{\beta_{\boldsymbol{\eta}}}{\sigma^2_{\boldsymbol{\eta}}} \right \} } I(\sigma^2_{\boldsymbol{\eta}}>0) \\
    &= (\sigma^2_{\boldsymbol{\eta}})^{-\alpha_{\boldsymbol{\eta}}^{*}-1} \exp{\left \{ -\frac{\beta_{\boldsymbol{\eta}}^{*}}{\sigma^2_{\boldsymbol{\eta}}} \right \} } I(\sigma^2_{\boldsymbol{\eta}}>0),
\end{align*}
 Hence, $\sigma^2_{\boldsymbol{\eta}}|\cdot \sim \hbox{IG} (\alpha_{\boldsymbol{\eta}}^{*},\beta_{\boldsymbol{\eta}}^{*}) I(\sigma^2_{\boldsymbol{\eta}}>0),$ where $\alpha_{\boldsymbol{\eta}}^{*}=\alpha_{\boldsymbol{\eta}}+\frac{r}{2}$, and $\beta_{\boldsymbol{\eta}}^{*}=\beta_{\boldsymbol{\eta}}+0.5(\boldsymbol{\eta}^{'}\boldsymbol{\eta})$.

\begin{align*}
    \sigma^2_{\boldsymbol{\xi}}|\cdot &\propto  (\sigma^2_{\boldsymbol{\xi}})^{-\frac{N}{2}}\exp{ \left \{ -\frac{1}{2} \boldsymbol{\xi}^{'}(\sigma^2_{\boldsymbol{\xi}}\mathbf{I}_r)^{-1}\boldsymbol{\xi}\right \} }\times (\sigma^2_{\boldsymbol{\xi}})^{-\alpha_{\boldsymbol{\xi}}-1} \exp{\left \{ -\frac{\beta_{\boldsymbol{\xi}}}{\sigma^2_{\boldsymbol{\xi}}} \right \} }I(\sigma^2_{\boldsymbol{\xi}}>0)\\
    &= (\sigma^2_{\boldsymbol{\xi}})^{-\alpha_{\boldsymbol{\xi}}^{*}-1} \exp{\left \{ -\frac{\beta_{\boldsymbol{\xi}}^{*}}{\sigma^2_{\boldsymbol{\xi}}} \right \} } I(\sigma^2_{\boldsymbol{\xi}}>0),
\end{align*}
Hence, $\sigma^2_{\boldsymbol{\xi}}|\cdot \sim \hbox{IG} (\alpha_{\boldsymbol{\xi}}^{*},\beta_{\boldsymbol{\xi}}^{*}) I(\sigma^2_{\boldsymbol{\xi}}>0),$ where $\alpha_{\boldsymbol{\xi}}^{*}=\alpha_{\boldsymbol{\xi}}+\frac{N}{2}$, and $\beta_{\boldsymbol{\xi}}^{*}=\beta_{\boldsymbol{\xi}}+0.5(\boldsymbol{\xi}^{'}\boldsymbol{\xi})$. 

\begin{align*}
    \tau^2|\cdot &\propto (\tau^2)^{-\frac{N}{2}}\exp{\left \{ -\frac{1}{2\tau^2} \left ( \mathbf{h}-\mathbf{W}\boldsymbol{\beta} -\mathbf{S}\boldsymbol{\delta}-\mathbf{M}\boldsymbol{\eta}-\boldsymbol{\xi}\right )^2 \right \}}\times (\tau^2)^{-\alpha_{\tau}-1} \exp{\left \{ -\frac{\beta_{\tau}}{\tau^2} \right \} } I(\tau^2>0)\\
    &= (\tau^2)^{-\alpha_{\tau}^{*}-1} \exp{\left \{ -\frac{\beta_{\tau}^{*}}{\tau^2} \right \} } I(\tau^2>0),
\end{align*}
Hence,
$\tau^2|\cdot \sim \hbox{IG} (\alpha_{\tau}^{*},\beta_{\tau}^{*}) I(\tau^2>0),$ where $\alpha_{\tau}^{*}=\alpha_{\tau}+\frac{N}{2}$, and \\ $\beta_{\tau}^{*}=\beta_{\tau}+0.5 \left ( \mathbf{h}-\mathbf{W}\boldsymbol{\beta} -\mathbf{S}\boldsymbol{\delta}-\mathbf{M}\boldsymbol{\eta}-\boldsymbol{\xi}\right )^2$.

\bibliographystyle{te}
\bibliography{reference}

\begin{thebibliography}{42}
\newcommand{\enquote}[1]{``#1''}
\providecommand{\natexlab}[1]{#1}
\providecommand{\url}[1]{\texttt{#1}}
\providecommand{\urlprefix}{URL }
\providecommand{\bibAnnoteFile}[1]{%
  \IfFileExists{#1}{\begin{quotation}\noindent\textsc{Key:} #1\\
  \textsc{Annotation:}\ \input{#1}\end{quotation}}{}}
\providecommand{\bibAnnote}[2]{%
  \begin{quotation}\noindent\textsc{Key:} #1\\
  \textsc{Annotation:}\ #2\end{quotation}}

\bibitem[{Arima et~al.(2017)Arima, Bell, Datta, Franco, and Liseo}]{Arima2017}
Arima, Serena, William~R Bell, Gauri~S Datta, Carolina Franco, and Brunero
  Liseo (2017), \enquote{Multivariate {Fay}--{Herriot} {Bayesian} estimation of
  small area means under functional measurement error.} \emph{Journal of the
  Royal Statistical Society: Series A (Statistics in Society)}, 180,
  1191--1209.
\bibAnnoteFile{Arima2017}

\bibitem[{Arima et~al.(2015)Arima, Datta, and Liseo}]{Arima2015}
Arima, Serena, Gauri~S. Datta, and Brunero Liseo (2015), \enquote{Bayesian
  estimators for small area models when auxiliary information is measured with
  error.} \emph{Scandinavian Journal of Statistics}, 42, 518--529.
\bibAnnoteFile{Arima2015}

\bibitem[{Banerjee et~al.(2015)Banerjee, Carlin, and
  Gelfand}]{BanerjeeGelfand2015}
Banerjee, Sudipto., Bradley~P. Carlin, and Alan~E. Gelfand (2015),
  \emph{Hierarchical Modeling and Analysis for Spatial Data}, 2 edition. CRC
  Press.
\bibAnnoteFile{BanerjeeGelfand2015}

\bibitem[{Bell et~al.(2007)Bell, Basel, Cruse, Dalzell, Maples, and
  Powers}]{Bell2007}
Bell, William~R., Wesley~W. Basel, Craig Cruse, Lucinda~P. Dalzell, Jerry~J.
  Maples, and David~S. Powers (2007), \enquote{Use of {A}{C}{S} data to produce
  {S}{A}{I}{P}{E} model-based estimates of poverty for counties.}
\bibAnnoteFile{Bell2007}

\bibitem[{Besag(1974)}]{Besag1974}
Besag, Julian (1974), \enquote{Spatial interaction and the statistical analysis
  of lattice systems.} \emph{Journal of the Royal Statistical Society: Series B
  (Methodological)}, 36, 192--225.
\bibAnnoteFile{Besag1974}

\bibitem[{Bradley et~al.(2011)Bradley, Cressie, and Shi}]{Bradley2011}
Bradley, Jonathan, Noel Cressie, and Tao Shi (2011), \enquote{Selection of rank
  and basis functions in the spatial random effects model.} volume 2011.
\bibAnnoteFile{Bradley2011}

\bibitem[{Bradley(2021)}]{Bradley2021}
Bradley, Jonathan~R. (2021), \enquote{Joint {B}ayesian analysis of multiple
  response-types using the {H}ierarchical {G}eneralized {T}ransformation
  model.} \emph{Bayesian Analysis}.
\bibAnnoteFile{Bradley2021}

\bibitem[{Bradley et~al.(2015)Bradley, Holan, and Wikle}]{BHW2015}
Bradley, Jonathan~R., Scott~H. Holan, and Christopher~K. Wikle (2015),
  \enquote{Multivariate spatio-temporal models for high-dimensional areal data
  with application to longitudinal employer household dynamics.} \emph{Annals
  of Applied Statistics}, 9, 1761--1791.
\bibAnnoteFile{BHW2015}

\bibitem[{Bradley et~al.(2018)Bradley, Holan, and Wikle}]{BHW2018}
Bradley, Jonathan~R., Scott~H. Holan, and Christopher~K. Wikle (2018),
  \enquote{Computationally efficient multivariate spatio-temporal models for
  high-dimensional count-valued data(with discussion).} \emph{Bayesian
  Analysis}, 13, 253--310.
\bibAnnoteFile{BHW2018}

\bibitem[{Bradley et~al.(2020)Bradley, Holan, and Wikle}]{Bradley2020}
Bradley, Jonathan~R., Scott~H. Holan, and Christopher~K. Wikle (2020),
  \enquote{Bayesian hierarchical models with conjugate full-conditional
  distributions for dependent data from the natural exponential family.}
  \emph{Journal of the American Statistical Association}, 115, 2037--2052.
\bibAnnoteFile{Bradley2020}

\bibitem[{Bradley et~al.(2017)Bradley, Wikle, and Holan}]{Bradley2017}
Bradley, Jonathan~R, Christopher~K Wikle, and Scott~H Holan (2017),
  \enquote{Regionalization of multiscale spatial processes by using a criterion
  for spatial aggregation error.} \emph{Journal of the Royal Statistical
  Society: Series B (Statistical Methodology)}, 79, 815--832.
\bibAnnoteFile{Bradley2017}

\bibitem[{Bradley et~al.(2019)Bradley, Wikle, and Holan}]{Bradley2019}
Bradley, Jonathan~R., Christopher~K. Wikle, and Scott~H. Holan (2019),
  \enquote{Spatio-temporal models for big multinomial data using the
  conditional multivariate logit-beta distribution.} \emph{Journal of Time
  Series Analysis}, 40, 363--382.
\bibAnnoteFile{Bradley2019}

\bibitem[{Cabral et~al.(2021)Cabral, de~Souza, and Leão}]{Cabral2021}
Cabral, Celso Rômulo~Barbosa, Nelson~Lima de~Souza, and Jeremias Leão (2021),
  \enquote{Bayesian measurement error models using finite mixtures of scale
  mixtures of skew-normal distributions.} \emph{Journal of Statistical
  Computation and Simulation}, 0, 1--22.
\bibAnnoteFile{Cabral2021}

\bibitem[{Carroll et~al.(2006)Carroll, Ruppert, Stefanski, and
  Crainiceanu}]{Caroll2006}
Carroll, Raymond~J., David. Ruppert, Leonard~A. Stefanski, and Ciprian~M.
  Crainiceanu (2006), \emph{Measurement Error in Nonlinear Models}, 2 edition.
  Chapman \& Hall/CRC.
\bibAnnoteFile{Caroll2006}

\bibitem[{Celeux et~al.(2006)Celeux, Forbes, Robert, and
  Titterington}]{Celeux2006}
Celeux, Gilles, Florence Forbes, Christian~P Robert, and D~Mike Titterington
  (2006), \enquote{Deviance information criteria for missing data models.}
  \emph{Bayesian analysis}, 1, 651--673.
\bibAnnoteFile{Celeux2006}

\bibitem[{Cressie and Wikle(2011)}]{Cressie2011}
Cressie, Noel and Christopher~K Wikle (2011), \emph{Statistics for
  spatio-temporal data}. John Wiley \& Sons.
\bibAnnoteFile{Cressie2011}

\bibitem[{Datta(2009)}]{Datta2009}
Datta, Gauri~S. (2009), \enquote{Chapter 32 - model-based approach to small
  area estimation.} In \emph{Handbook of Statistics} (C.R. Rao, ed.), volume~29
  of \emph{Handbook of Statistics}, 251--288, Elsevier.
\bibAnnoteFile{Datta2009}

\bibitem[{Diaconis and Ylvisaker(1979)}]{Diaconis1979}
Diaconis, Persi and Donald Ylvisaker (1979), \enquote{Conjugate priors for
  exponential families.} \emph{Ann. Statist.}, 7, 269--281.
\bibAnnoteFile{Diaconis1979}

\bibitem[{Fay and Herriot(1979)}]{FH1979}
Fay, Robert~E. and Roger~A. Herriot (1979), \enquote{Estimates of income for
  small places: An application of james-stein procedures to census data.}
  \emph{Journal of the American Statistical Association}, 74, 269--277.
\bibAnnoteFile{FH1979}

\bibitem[{Fuller(2009)}]{Fuller2009}
Fuller, Wayne~A (2009), \emph{Measurement error models}, volume 305. John Wiley
  \& Sons.
\bibAnnoteFile{Fuller2009}

\bibitem[{Gelfand and Vounatsou(2003)}]{Gelfand2003}
Gelfand, Alan~E and Penelope Vounatsou (2003), \enquote{Proper multivariate
  conditional autoregressive models for spatial data analysis.}
  \emph{Biostatistics}, 4, 11--15.
\bibAnnoteFile{Gelfand2003}

\bibitem[{Gelman et~al.(2014)Gelman, Hwang, and Vehtari}]{Gelman2014}
Gelman, Andrew, Jessica Hwang, and Aki Vehtari (2014), \enquote{Understanding
  predictive information criteria for bayesian models.} \emph{Statistics and
  computing}, 24, 997--1016.
\bibAnnoteFile{Gelman2014}

\bibitem[{Griffith(2000)}]{Grif2000}
Griffith, Daniel~A. (2000), \enquote{A linear regression solution to the
  spatial autocorrelation problem.} \emph{Journal of Geographical Systems}, 2,
  141--156.
\bibAnnoteFile{Grif2000}

\bibitem[{Hoben and Richardson(1992)}]{Hoben1992}
Hoben, James and Todd Richardson (1992), \enquote{The local {C}{H}{A}{S}: {A}
  preliminary assessment of first year submissions.} \emph{Office of Policy
  Development and Research, U.S. Department of Housing and Urban Development}.
\bibAnnoteFile{Hoben1992}

\bibitem[{Hughes and Haran(2013)}]{HH2013}
Hughes, John and Murali Haran (2013), \enquote{Dimension reduction and
  alleviation of confounding for spatial generalized linear model.}
  \emph{Journal of the Royal Statistical Society: Series B (Statistical
  Methodology)}, 75, 139--159.
\bibAnnoteFile{HH2013}

\bibitem[{Huques et~al.(2014)Huques, Bondell, and Ryan}]{Huque2014}
Huques, Md~Hamidul, Howard Bondell, and Louise Ryan (2014), \enquote{On the
  impact of covariate measurement error on spatial regression modelling.}
  \emph{Environmetrics}, 25, 560--570.
\bibAnnoteFile{Huque2014}

\bibitem[{Jin et~al.(2007)Jin, Banerjee, and Carlin}]{Jin2007}
Jin, Xiaoping, Sudipto Banerjee, and Bradley~P Carlin (2007),
  \enquote{Order-free co-regionalized areal data models with application to
  multiple-disease mapping.} \emph{Journal of the Royal Statistical Society:
  Series B (Statistical Methodology)}, 69, 817--838.
\bibAnnoteFile{Jin2007}

\bibitem[{Jin et~al.(2005)Jin, Carlin, and Banerjee}]{Jin2005}
Jin, Xiaoping, Bradley~P Carlin, and Sudipto Banerjee (2005),
  \enquote{Generalized hierarchical multivariate car models for areal data.}
  \emph{Biometrics}, 61, 950--961.
\bibAnnoteFile{Jin2005}

\bibitem[{Li et~al.(2009)Li, Tang, and Lin}]{Li2009}
Li, Yi, Haicheng Tang, and Xihong Lin (2009), \enquote{Spatial linear mixed
  models with covariate measurement errors.} \emph{Statistica Sinica}, 19,
  1077--1093.
\bibAnnoteFile{Li2009}

\bibitem[{Matos et~al.(2018)Matos, Castro, Cabral, and Lachos}]{Matos2018}
Matos, Larissa~A, Luis~M Castro, Celso~RB Cabral, and V{\'\i}ctor~H Lachos
  (2018), \enquote{Multivariate measurement error models based on student-t
  distribution under censored responses.} \emph{Statistics}, 52, 1395--1416.
\bibAnnoteFile{Matos2018}

\bibitem[{Obled and Creutin(1986)}]{Obled1986}
Obled, Ch and JD~Creutin (1986), \enquote{Some developments in the use of
  empirical orthogonal functions for mapping meteorological fields.}
  \emph{Journal of Applied Meteorology and Climatology}, 25, 1189--1204.
\bibAnnoteFile{Obled1986}

\bibitem[{Ogden and Tarpey(2006)}]{Ogden2006}
Ogden, R. and Thaddeus Tarpey (2006), \enquote{Estimation in regression models
  with externally estimated parameters.} \emph{Biostatistics (Oxford,
  England)}, 7, 115--29.
\bibAnnoteFile{Ogden2006}

\bibitem[{Pfeffermann(2013)}]{Pfeffermann2013}
Pfeffermann, Danny (2013), \enquote{New important developments in small area
  estimation.} \emph{Statistical Science}, 28, 40--68.
\bibAnnoteFile{Pfeffermann2013}

\bibitem[{Porter and Oleson(2014)}]{Porter2014a}
Porter, Aaron~T. and Jacob~J. Oleson (2014), \enquote{A multivariate car model
  for mismatched lattices.} \emph{Spatial and Spatio-temporal Epidemiology},
  11, 79--88.
\bibAnnoteFile{Porter2014a}

\bibitem[{Rao and Molina(2015)}]{Rao2015}
Rao, John~NK and Isabel Molina (2015), \emph{Small area estimation}. John Wiley
  \& Sons.
\bibAnnoteFile{Rao2015}

\bibitem[{Reich et~al.(2006)Reich, Hodges, and Zadnik}]{Reich2006}
Reich, Brian~J, James~S Hodges, and Vesna Zadnik (2006), \enquote{Effects of
  residual smoothing on the posterior of the fixed effects in disease-mapping
  models.} \emph{Biometrics}, 62, 1197--1206.
\bibAnnoteFile{Reich2006}

\bibitem[{Spiegelhalter et~al.(2002)Spiegelhalter, Best, Carlin, and Van
  Der~Linde}]{Spiegelhalter2002}
Spiegelhalter, David~J, Nicola~G Best, Bradley~P Carlin, and Angelika Van
  Der~Linde (2002), \enquote{Bayesian measures of model complexity and fit.}
  \emph{Journal of the royal statistical society: Series b (statistical
  methodology)}, 64, 583--639.
\bibAnnoteFile{Spiegelhalter2002}

\bibitem[{Tadayon and Torabi(2019)}]{Tadayon2018}
Tadayon, Vahid and Mahmoud Torabi (2019), \enquote{Spatial models for
  non-gaussian data with covariate measurement error.} \emph{Environmetrics},
  30, e2545.
\bibAnnoteFile{Tadayon2018}

\bibitem[{USCB(2020)}]{Census2020}
USCB (2020), \enquote{Understanding and using american community survey data:
  What all data users need to know.}
\bibAnnoteFile{Census2020}

\bibitem[{Wall(2004)}]{Wall2004}
Wall, Melanie~M (2004), \enquote{A close look at the spatial structure implied
  by the car and sar models.} \emph{Journal of statistical planning and
  inference}, 121, 311--324.
\bibAnnoteFile{Wall2004}

\bibitem[{Watanabe and Opper(2010)}]{Watanabe2010}
Watanabe, Sumio and Manfred Opper (2010), \enquote{Asymptotic equivalence of
  bayes cross validation and widely applicable information criterion in
  singular learning theory.} \emph{Journal of machine learning research}, 11.
\bibAnnoteFile{Watanabe2010}

\bibitem[{Ybarra and Lohr(2008)}]{YL2008}
Ybarra, Lynn M.~R. and Sharon~L. Lohr (2008), \enquote{Small area estimation
  when auxiliary information is measured with error.} \emph{Biometrika}, 95,
  919--931.
\bibAnnoteFile{YL2008}

\end{thebibliography}

\end{document}